\documentclass[amsmath,amssymb,onecolumn,superscriptaddress]{revtex4}

\usepackage{graphicx}
\usepackage{dcolumn}

\def\be{\begin{equation}}
\def\ee{\end{equation}}
\def\beq{\begin{eqnarray}}
\def\eeq{\end{eqnarray}}

\usepackage{bm}
\usepackage{graphicx}
\usepackage{dcolumn}
\usepackage{amsmath}
\usepackage[latin1]{inputenc}
\usepackage{graphicx, psfrag}
\usepackage{amssymb}
\usepackage[colorlinks=true, citecolor=blue, urlcolor = blue, linkcolor= red, bookmarks=true]{hyperref}
\usepackage{float}
\usepackage{amsmath}
\usepackage{tensor}
\usepackage{amsfonts}
\usepackage{dcolumn}
\usepackage{hyperref}
\usepackage{subfigure}
\usepackage{pgfplots}
\usepackage{epstopdf}
\usepackage{booktabs}

\begin{document}

    \title{\bf{Constraints on the parameters of modified Chaplygin-Jacobi and modified Chaplygin-Abel gases in $f(T)$ gravity}}
    \author{Himanshu Chaudhary}
    \email{himanshuch1729@gmail.com}
    \affiliation{Pacif Institute of Cosmology and Selfology (PICS), Sagara, Sambalpur 768224, Odisha, India}
    \affiliation{Department of Applied Mathematics, Delhi Technological University, Delhi-110042, India,}
     \affiliation{Department of Mathematics, Shyamlal College, University of Delhi, Delhi-110032, India,}
    \author{Ujjal Debnath}
    \email{ujjaldebnath@gmail.com}
    \author{Tanusree Roy}
    \email{tanusreeroy1995@gmail.com} \affiliation{Department of
        Mathematics, Indian Institute of Engineering Science and
        Technology,\\ Shibpur, Howrah-711 103, India,}
   \author{Sayani Maity}
   \email{sayani.maity88@gmail.com}
   \affiliation{Department of Mathematics, Sister Nivedita University, DG-1/2, Action Area 1, New Town, Kolkata-700 156, India.}
\author{G.Mustafa}
\email{gmustafa3828@gmail.com} 
\affiliation{Department of Physics,
Zhejiang Normal University, Jinhua 321004, Peoples Republic of China,}
\affiliation{New Uzbekistan University, Mustaqillik ave. 54, 100007 Tashkent, Uzbekistan,}
\author{Monika Arora}
\email{monika.arora@mirandahouse.ac.in} 
\affiliation{Department of Mathematics, Miranda House, University of Delhi, Delhi-110007,}

\begin{abstract}
In this work, we explore the parameter constraints of two dark energy models, namely the Modified Chaplygin-Jacobi gas (MCJG) and Modified Chaplygin-Abel gas (MCAG), within the context of $f(T)$ gravity model in a non-flat Friedmann-Lema\^itre-Robertson-Walker (FLRW) Universe. Our investigation involves comparing the equation of state for the MCJG and MCAG dark energy models with the equation of state derived from the $f(T)$ gravity model. To derive constraints for the dark energy and $f(T)$ gravity models, we use recent astronomical datasets, including $H(z)$ data, type Ia supernovae observations, Gamma Ray Bursts data, quasar data, and Baryon Acoustic Oscillation (BAO) measurements. We present the reduced Hubble parameter in terms of observable parameters such as $\Omega_{r0}$ (density parameter of radiation), $\Omega_{m0}$ (density parameter of dark matter), $\Omega_{k0}$ (density parameter associated with spatial curvature), $\Omega_{CJ0}$ (density parameter of Modified Chaplygin-Jacobi gas), $\Omega_{CA0}$ (density parameter of Modified Chaplygin-Abel gas), and $H_0$ (present value of the Hubble parameter). We explore the cosmological evolution through various cosmic diagnostic parameters, including the deceleration parameter, $Om(z)$ diagnostic, and statefinder diagnostic pair $\{r,s\}$. These diagnostic parameters offer valuable insights into the expansion dynamics and the nature of dark energy in the Universe. We have also assessed the viability of the models using the information criteria. Our aim is to shed light on the nature of dark energy and its connection to the $f(T)$ gravity model, and ultimately gain a deeper understanding of the underlying mechanisms driving the accelerated expansion of our Universe.
\end{abstract}

\date{\today}

\maketitle

    \section{Introduction}\label{I}
    Over the course of last decade numerous astronomical observations have been carried out to
    investigate the compatibility of theoretical cosmological models.
    Recent observations of the type Ia supernova
    (SNIa)\cite{1:1997zqe,2:1998vns,3:1998fmf}, Baryon Acoustic
    Oscillation (BAO)\cite{4:1997ik,5:2005xqv}, cosmic microwave
    background (CMB)\cite{6:2003yz}, WMAP \cite{7:2003elm,8:2008lyn}
    have confirmed the most astounding result of modern cosmology that not only our Universe is expanding but it is passing through an era of accelerated expansion. Now, the standard Big Bang cosmology was unable to resolve
    this `cosmic puzzle'; so the possible existence of some exotic
    matter, acclaimed as Dark Energy (DE), with a unique characteristic of negative pressure, has emerged which would obey $\rho+3p<0$ as the
    Friedmann equation reads

    \begin{equation}
        \frac{\ddot{a}}{a}=-\frac{4\pi G}{3}(\rho+3p)
    \end{equation}

    and $\ddot{a}>0$ for accelerated expansion. At first $\Lambda$CDM
    model was considered as the `best fit'\cite{9:2002gy,10}, but fine-tuning problem and the coincidence problem could not be explained with this model \cite{11:1988cp,12,13:2006wr}. To resolve these severe issues theorists suggested for modification of Einsteins field equations in two ways. One approach is the correction of the matter field that yields various dynamical dark energy models with varying equation of state such as quintessence, tachyon,
 k-essence, fermionic field, phantom, Chaplygin gas, DBI essenec\cite{13:2006wr,14} etc. In search of a befitting candidate for DE, Chaplygin\cite{15} originated a model called Chaplygin gas (CG), defined with a simple equation of state

    \begin{equation}
        p=-\frac{B}{\rho}
    \end{equation}

    where $B$ is considered as positive constant. This model unified dark matter
    and DE because it acts as dust-like matter at early stage and cosmological constant at later time. But one backdrop of this model is that it can not explain the era of the structure formation in the Universe \cite{16A,16B}. In this regard, the pure CG model is generalized to the generalized Chaplygin gas (GCG) model \cite{16C} with  equation of state

    \begin{equation}
        p=-\frac{B}{\rho^\alpha},~~\mbox{with} ~~0<\alpha\leq1.
    \end{equation}
   In Ref.\cite{16,17} Benaoum proposed
    Modified Chaplygin Gas (MCG) model that obeys the EoS

    \begin{equation}
        p=A\rho-\frac{B}{\rho^{\alpha}}
    \end{equation}

    where  $A,B>0$ and $0\leq\alpha\leq1$. For $A=1/3$, the model exhibits a radiation era when assuming a minuscule scale factor. On the other hand, with an infinitely large scale factor, it acts as the $\Lambda$CDM model. In an intermediate stage, it behaves as pressure-less dust. Further along, the EoS was revised to derive other extensions like variable CG\cite{18}, variable modified CG\cite{19:2007bw},  generalized cosmic Chaplygin gas (GCCG), the new variable modified Chaplygin gas (NVMCG)\cite{22A},
 viscous CG\cite{20:2005mu} etc. For better synchronization of the models to the observational data, constraints on different parameters of CG and MCG were investigated\cite{21,22,23:2014ura,24:2014kza,25:2013ara,26}. Using Jacobi's elliptic function, Villanueva formulated generalized Chaplygin-Jacobi gas correlated with the generalized Chaplygin scalar field\cite{27,28,29} 
 which can describe the inflationary stage and confronted the measurement recently released by the Planck 2015 data.
Following that, Debnath derived the equation of states for Modified Chaplygin-Abel gas and Modified Chaplygin-Jacobi gas and explored the models' compatibility with observational data using AIC, BIC, and DIC model selection tools\cite{30:2021pxy}. Accretion phenomena of Modified Chaplygin-Jacobi Gas and Modified Chaplygin-Abel Gas models into Schwarzschild Black Hole have been studied in ref \cite{30A} .\\

     Another approach to explain the cosmic acceleration of the Universe
    is constructing gravitational modifications in the geometric part of the Einsteins field equations so that the additional
    degree(s) of freedom can act as a driving force in the accelerated expansion
    \cite{31:2011et}. Mostly modified gravity theories are formed
    by extending the Einstein-Hilbert action i.e. by curvature
    formulation. Some of the widely used examples are $f(R)$ gravity
    \cite{32:2010wj,33:2010aj}; $f(G)$ gravity\cite{34:2008wz};
    Horava-Lifshitz gravity\cite{35}; Lovelock gravity\cite{36:1971yv}
    etc. Another suitable approach to gravitational modification is
    the extension of the action of equivalent torsional formulation of
    General Relativity (GR). The simplest modification is $f(T)$
    gravity \cite{37:2008ey,38:2008gz,39:2010py}, developed from the
    Teleparallel Equivalent of General Relativity (TEGR)
    \cite{40:2005in,41,42,43:1979qx,44,45}, where $T$ is the Torsion
    scalar. The characteristic difference between TEGR with GR is that
    the Lagrangian of TEGR contains a curvature-less Weitzenbock
    connection whereas GR has a torsion-less  Levi-Civita connection.
    The desirability of $f(T)$ gravity model over $f(R)$
    gravity is the second-order field equation in contrast with fourth-order field equations in $f(R)$ gravity. In teleparallelism, torsion is formed from the product of
    the first derivatives of the tetrad having no second derivative
    terms. Thus the Lagrangian can simply be extended by just adding a
    function $f(T)$. Both the inflation phase
    \cite{47:2006jd,48:2014zra,49:2016gbu,50} and acceleration
    phase\cite{51:2010va,52:2010xk,53} of the Universe have been
    studied under the framework of $f(T)$ gravity. As $f(T)$ gravity
    incorporates an unknown function, the viability of the model
    remains questionable from theoretical and observational
    perspectives depending on the function chosen. Using Monte-Carlo
    analysis, $f(T)$ gravity models have been investigated
    already\cite{54}. Constraining the free parameters have been
    executed with data sources like Pantheon Supernovae
    sample\cite{55:2019miu}, redshift-space distribution measurements
    \cite{55:2019miu} or solar system data\cite{56:2012cm}. Previous
    works in this domain have suggested that DE coupled with a
    modified gravity field can present fascinating results. For
    example, Jamil et al answered the cosmic coincidence problem by
    working on MCG in loop quantum gravity\cite{57:2011rh}. Park
    pointed out that additional terms like $a^{-4}, a^{-2}$ contained
    in Friedmann equation in Horava-Lifshitz gravity can be
    responsible for cosmic acceleration\cite{58}. Many other gravity
    theories like Chern-Simons gravity\cite{59:2014saa}, Galileon
    gravity\cite{60:2014fra}, Einstein-Aether
    gravity\cite{61:2013cja}, Brans-Dicke theory\cite{62},
    were combined with MCG to determine the observational constraints
    on the parameters. Motivated by this line of work, here we will
    review the observational constraints of MCJG and MCAG parameters
    with association of various $f(T)$ gravity models. The free
    parameters of the $f(T)$ models have also been constrained with the help of different
    combinations of current observational data.\\

The flow of work of this paper is as follows:
the basic equations of $f(T)$ gravity models is described in section \ref{secII}. In section \ref{secIII} the viable $f(T)$ gravity model  is presented. Section \ref{secIV} is devoted to investigate the constraints of the model parameters using various observational dataset comprising of $H(z)$, Pantheon, Gamma Ray Bursts, quasars, and BAO measurements. Comparison of observational and theoretical predictions of model parameters are done via $H(z)$ measurements and distance modulus $\mu(z)$ in section \ref{secV}. For cosmographic analysis the deceleration parameter, jerk parameter are derived in section \ref{secVI}. The statefinder diagnostic $Om(z)$ diagnostic are analyzed graphically in section \ref{secVII} and \ref{secVIII} respectively. In section \ref{sec9}, we present the information criteria for both the models. Finally in section \ref{secIX}, the results are summarized and the main conclusion is exposed.

\section{Basic Equations in $f(T)$ Gravity}\label{secII}
One of the most intriguing generalizations of Einstein general relativity in the context of explaining the observed cosmic acceleration of the Universe is the $f(T)$ theory of gravity. Originally, teleparallelism was introduced by Einstein in the process of unifying gravity and electromagnetism. In contrast to the General Theory of Relativity, which uses the torsionless Levi-Civita connection, this theory exploits the curvatureless Weitzenbock connection that possesses only torsion. The geometry of a non-flat Friedmann-Lema\^itre-Robertson-Walker (FLRW) model of the Universe is described by the metric. 
\begin{eqnarray}
        ds^2=-dt^2 +a^2(t)\left[\frac{dr^2}{1-kr^{2}}+r^2\left(d\theta^2 +
        sin^2 \theta d\phi^2 \right)\right]
\end{eqnarray}

where $a(t)$ is the scale factor and $k(=0,-1,+1)$ is the curvature scalar which represents flat ($k=0$), open ($k=1$), and closed ($k=-1$) model of the Universe.  The relation of orthonormal tetrad components $\bf{e}_A(x^{\mu})$ relate and the metric is 
\begin{equation}
    g_{\mu \nu}=\eta_{AB}e_\mu ^A(x)e_\nu ^B(x),
\end{equation}
where $\bf{e}_A(x^{\mu})=e_\mu ^A \partial_\mu$, the indices $A, B$ run over 0, 1, 2, 3 for the tangent space of the manifold, the coordinate indices on the manifold $\mu, \nu$ take values 0, 1, 2, 3 and $i, j$ are the spatial indices. In this formalism the torsion tensor takes the form
\begin{equation}
    T^{\lambda}_{\mu\nu}=\Gamma^{\bf{W}^\lambda}_{\nu\mu}-\Gamma^{\bf{W}^\lambda}_{\mu\nu}=e_A^\lambda(\partial_\mu e_\nu ^A -\partial_\nu e_\mu ^A).
\end{equation}
The contorsion tensor reads
\begin{equation}
    K^{\mu\nu}_{\rho}=-\frac{1}{2}(T^{\mu\nu}_{\rho}-T^{\nu\mu}_{\rho}-T^{\mu\nu}_{\rho})
\end{equation}
From torsion tensor and contorsion tensor a new quantity is defined
\begin{equation}
    S^{\mu\nu}_{\rho}=\frac{1}{2}(K^{\mu\nu}_{\rho}+\delta_\rho^{\mu} T^{\alpha\nu}_{\alpha}-\delta_\rho^{\nu}T^{\alpha\mu}_{\alpha})
\end{equation}
The Lagrangian of the Telepaarallel gravity is the torsion scalar $T$ defined as follows
\begin{equation}
    T=S^{\mu\nu}_{\rho}T^{\rho}_{\mu\nu}.
\end{equation}

Following the footsteps of $f(R)$ gravity, $f(T)$ gravity is formed by extending the torsion scalar $T$ to $T+f(T)$ where $f(T)$ is a
    differentiable function of $T$. The action for $f(T)$ gravity can
    be written as\cite{38:2008gz,39:2010py,63:2011tc,64}

    \begin{equation}
        S=\frac{1}{16\pi G}\int d^4 x \sqrt{-g}\left[T+f(T)+{\cal{L}}_m
        \right]
    \end{equation}

    where ${\cal{L}}_m$ is denoted as the Lagrangian for matter content. TEGR model is retrieved as $f(T)\rightarrow 0$. It exhibits the $\Lambda$CDM model as $f(T)$ approaches a constant value $\Lambda$.\\

    In $f(T)$ gravity, the Friedmann equations are written for non-flat FLRW Universe  \cite{65,66}

    \begin{equation}\label{F1}
        \left(H^{2}
        +\frac{k}{a^{2}}\right)+2H^{2}f'(T)+\frac{f(T)}{6}=\frac{8\pi
            G}{3}~ \rho
    \end{equation}

    \begin{equation}\label{F2}
        \left(\dot{H} -\frac{k}{a^{2}}\right)\left(1+f'(T)+12H^{2}f''(T)
        \right)=-4\pi G(\rho +p)
    \end{equation}

    where, $H=\dot{a}/a$ is the Hubble parameter and the torsion
    scalar $T=-6\left(H^{2}+\frac{k}{a^{2}}\right)$ with prime and dot
    denote the derivatives w.r.t. $T$ and $t$ respectively. We assume
    that the total energy density and pressure of the fluid that fills
    the Universe are $\rho$ and $p$, respectively. So the energy
    conservation equation takes the form

    \begin{equation}\label{Cons}
        \dot{\rho}+3H(\rho+p)=0
    \end{equation}

    We start with the prediction that the Universe is composed of
    matter content comprising radiation, dark matter (DM), and dark
    energy (DE). So, $\rho$ and $p$ consist of densities and pressures
    of radiation, DM and DE. So $\rho=\rho_{r}+\rho_{m}+\rho_{d}$ and
    $p=p_{r}+p_{m}+p_{d}$. Now assume that radiation, DM, and DE
    follow the conservation equation separately so that

    \begin{equation}\label{rad}
        \dot{\rho}_{r}+3H(\rho_{r}+p_{r})=0,
    \end{equation}

    \begin{equation}\label{DM}
        \dot{\rho}_{m}+3H(\rho_{m}+p_{m})=0
    \end{equation}

    and

    \begin{equation}\label{DE}
        \dot{\rho}_{d}+3H(\rho_{d}+p_{d})=0
    \end{equation}

    The relation between energy density and pressure for radiation is,
    $p_{r}=\frac{1}{3}\rho_{r}$. Solving equation (\ref{rad}) we
    obtain $\rho_{r}=\rho_{r0}a^{-4}$. Since pressure for the DM can
    be considered negligible  (i.e., $p_{m}=0$), equation
    (\ref{DM}) gives $\rho_{m}=\rho_{m0}a^{-3}$.\\
    $\bullet{}$ {\bf MCJG :} Now consider the DE in the form of
    modified Chaplygin-Jacobi gas (MCJG) whose equation of state is
    given by \cite{67}

    \begin{equation}\label{MCJG}
        p_{d}=[(2K-1)(1+A)-1]\rho_{d}-\frac{KB}{\rho_{d}^{\alpha}}+\frac{(1-K)(1+A)^{2}}{B}~\rho_{d}^{2+\alpha}
    \end{equation}

    where $K\in[0,1]$, $\alpha\in[0,1]$, $A(>0)$ and $B(>0)$ are
    constants. Putting the above expression in (\ref{DE}), we obtain

    \begin{equation}\label{rhoMCJG}
        \rho_{d}^{1+\alpha}=\frac{B}{1+A}\left[\frac{A_{s}+(1-A_{s})(1+z)^{3(1+\alpha)(1+A)}}
        {KA_{s}-(1-K)(1-A_{s})(1+z)^{3(1+\alpha)(1+A)}} \right]
    \end{equation}

    where $A_{s}$ is constant satisfying $1-K<A_{s}<1$. So the present
    value of the energy density is
    $\rho_{CJ0}^{1+\alpha}=\frac{B}{(1+A)[KA_{s}-(1-K)(1-A_{s})]}$.\\
    If the DE of the Universe obeys the MCJG equation of state
    (\ref{MCJG}), then using the equation (\ref{rhoMCJG}), the
    equation (\ref{F1}) reduces to the following equation for $E(z)$
    as

    \begin{eqnarray}\label{MCJG1}
        E^{2}(z)=\Omega_{r0}(1+z)^{4}+\Omega_{m0}(1+z)^{3}-\Omega_{k0}(1+z)^{2}
        +\Omega_{CJ0}[KA_{s}-(1-K)(1-A_{s})]^{\frac{1}{1+\alpha}}
        \nonumber\\
        \times\left[\frac{A_{s}+(1-A_{s})(1+z)^{3(1+\alpha)(1+A)}}
        {KA_{s}-(1-K)(1-A_{s})(1+z)^{3(1+\alpha)(1+A)}}
        \right]^{\frac{1}{1+\alpha}}  - 2f'(T)-\frac{f(T)}{6H_{0}^{2}}
    \end{eqnarray}

    where we have defined the normalized Hubble parameter
    $E(z)=\frac{H(z)}{H_{0}}$ and the dimensionless density parameters
    $\Omega_{r0}=\frac{8\pi G\rho_{r0}}{3H_0^2}$,
    $\Omega_{m0}=\frac{8\pi G\rho_{m0}}{3H_0^2}$,
    $\Omega_{k0}=\frac{k}{H_0^2}$ and $\Omega_{CJ0}=\frac{8\pi
        G\rho_{_{CJ0}}}{3H_0^2}$. Putting $z=0$ in the above equation, we
    obtain $\Omega_{r0}+\Omega_{m0}-\Omega_{k0}+\Omega_{CJ0}-
    2f'(T_{0})-\frac{f(T_{0})}{6H_{0}^{2}}=1$
    where $T_{0}=-6H_{0}^{2}\left(1+\Omega_{k0}\right)$.\\
    $\bullet{}$ {\bf MCAG :} Next, we consider the DE in the form of
    modified Chaplygin-Abel gas (MCJG) whose equation of state is
    given by \cite{67}

    \begin{equation}\label{MCAG}
        p_{d}=[(b^{2}+2c^{2})(1+A)-1]\rho_{d}-\frac{c^{2}B}{\rho_{d}^{\alpha}}-\frac{(b^{2}+c^{2})(1+A)^{2}}{B}~\rho_{d}^{2+\alpha}
    \end{equation}

    where $b,c\in \mathbb{R}$. Putting the above expression in
    (\ref{DE}), we obtain

    \begin{equation}\label{rhoMCAG}
        \rho_{d}^{1+\alpha}=\frac{c^{2}B}{1+A}\left[\frac{B_{s}+(1-B_{s})(1+z)^{3b^{2}(1+\alpha)(1+A)}}
        {c^{2}B_{s}+(b^{2}+c^{2})(1-B_{s})(1+z)^{3b^{2}(1+\alpha)(1+A)}}
        \right]
    \end{equation}

    where $B_{s}$ is a constant satisfying $0<B_{s}<1$. So the present
    value of the energy density is
    $\rho_{CA0}^{1+\alpha}=\frac{c^{2}B}{(1+A)[c^{2}B_{s}+(b^{2}+c^{2})(1-B_{s})]}$.\\

    If the DE of the Universe obeys the MCAG equation of state
    (\ref{MCAG}), then using the equation (\ref{rhoMCAG}), the
    equation (\ref{F1}) reduces to the following equation for $E(z)$
    as

    \begin{eqnarray}\label{MCAG1}
        E^{2}(z)=\Omega_{r0}(1+z)^{4}+\Omega_{m0}(1+z)^{3}-\Omega_{k0}(1+z)^{2}
        +\Omega_{CA0}[c^{2}B_{s}+(b^{2}+c^{2})(1-B_{s})]^{\frac{1}{1+\alpha}}
        \nonumber\\
        \times\left[\frac{B_{s}+(1-B_{s})(1+z)^{3b^{2}(1+\alpha)(1+A)}}
        {c^{2}B_{s}+(b^{2}+c^{2})(1-B_{s})(1+z)^{3b^{2}(1+\alpha)(1+A)}}
        \right]^{\frac{1}{1+\alpha}} - 2f'(T)-\frac{f(T)}{6H_{0}^{2}}
    \end{eqnarray}
    where the dimensionless density parameter $\Omega_{CA0}=\frac{8\pi
        G\rho_{_{CA0}}}{3H_0^2}$. Putting $z=0$ in the above equation, we
    obtain $\Omega_{r0}+\Omega_{m0}-\Omega_{k0}+\Omega_{CA0}-
    2f'(T_{0})-\frac{f(T_{0})}{6H_{0}^{2}}=1$.\\

    \section{Viable $f(T)$ Model}\label{secIII}
    Theoretically, the function $f(T)$ can be elected independently;
    but in practice it should be chosen in accordance with the fact
    the divergence from TEGR should be very small. Keeping that in
    mind, various feasible models were proposed over the years.
    Bengochea and Ferraro have introduced the power-law model
    characterized by \cite{38:2008gz}:

    \begin{equation}
        f(T)=\beta (-T)^{\gamma}
    \end{equation}
    where $\beta$ and $\gamma$ are the two model parameters. For
    $\beta=0$, spatially flat Friedmann equation is recovered. To
    attain accelerated expansion from this model, the restriction
    $\gamma<1$ need to be imposed.\\

    $\bullet{}$ {\bf MCJG:} From equation (\ref{MCJG1}), we obtain
    \begin{eqnarray}
        E^{2}(z)+6\beta
        H_{0}^{2}\left[E^{2}(z)+\Omega_{k0}(1+z)^{2}-2\gamma
        \right]\left[E^{2}(z)+\Omega_{k0}(1+z)^{2}\right]^{\gamma-1}
        \nonumber\\
        =\Omega_{r0}(1+z)^{4}+\Omega_{m0}(1+z)^{3}-\Omega_{k0}(1+z)^{2}
        +\Omega_{CJ0}[KA_{s}-(1-K)(1-A_{s})]^{\frac{1}{1+\alpha}}
        \nonumber\\
        \times\left[\frac{A_{s}+(1-A_{s})(1+z)^{3(1+\alpha)(1+A)}}
        {KA_{s}-(1-K)(1-A_{s})(1+z)^{3(1+\alpha)(1+A)}}
        \right]^{\frac{1}{1+\alpha}}
    \end{eqnarray}
    with $\Omega_{r0}+\Omega_{m0}-\Omega_{k0}+\Omega_{CJ0}=1+6\beta
    H_{0}^{2}(1+\Omega_{k0}-2\gamma)(1+\Omega_{k0})^{\gamma-1} $.\\

    $\bullet{}$ {\bf MCAG:} From equation (\ref{MCAG1}), we obtain
    \begin{eqnarray}
        E^{2}(z)+6\beta
        H_{0}^{2}\left[E^{2}(z)+\Omega_{k0}(1+z)^{2}-2\gamma
        \right]\left[E^{2}(z)+\Omega_{k0}(1+z)^{2}\right]^{\gamma-1}
        \nonumber\\
        =\Omega_{r0}(1+z)^{4}+\Omega_{m0}(1+z)^{3}-\Omega_{k0}(1+z)^{2}
        +\Omega_{CA0}[c^{2}B_{s}+(b^{2}+c^{2})(1-B_{s})]^{\frac{1}{1+\alpha}}
        \nonumber\\
        \times\left[\frac{B_{s}+(1-B_{s})(1+z)^{3b^{2}(1+\alpha)(1+A)}}
        {c^{2}B_{s}+(b^{2}+c^{2})(1-B_{s})(1+z)^{3b^{2}(1+\alpha)(1+A)}}
        \right]^{\frac{1}{1+\alpha}}
    \end{eqnarray}
    with $\Omega_{r0}+\Omega_{m0}-\Omega_{k0}+\Omega_{CA0}=1+6\beta
    H_{0}^{2}(1+\Omega_{k0}-2\gamma)(1+\Omega_{k0})^{\gamma-1} $.\\

\section{Data Analysis}\label{secIV}
In this section, we will perform a comprehensive comparison between the predictions of the Modified Chaplygin-Abel gas (MCAG) and Modified Chaplygin-Jacobi gas (MCJG) models with observational data. Our objective is to understand the constraints placed on these models by utilizing different distinct datasets: the $H(z)$ dataset, the ype Ia supernova, Gamma Ray Bursts (GRBs), quasa (Q) , Baryon Acoustic Oscillation (BAO). By comparing the model predictions with these cosmological data sets, We aim to determine the best-fit values for free parameters of each model. In our analysis, we will also incorporate the present-day value of the Hubble function $H_{0}$ to provide a robust estimation of the model parameters. To achieve this, we adopt a rigorous and widely-used approach based on Bayesian statistics. This statistical technique involves the use of likelihood functions to quantify the agreement between the model predictions and the observational data. By applying the Markov Chain Monte Carlo (MCMC) method within the Bayesian framework, we can efficiently explore the parameter space and sample the probability distributions of each parameters. The Bayesian approach offers several advantages, including the ability to include prior information or constraints from previous studies and to handle uncertainties in the model and observational data. By employing this method, we can obtain robust constraints on the parameters of the MCAG and MCJG models and gain valuable insights into their viability in explaining the observed cosmological phenomena. The statistical analysis will allow us to determine the regions in parameter space where the models are consistent with the data and, consequently, provide valuable guidance for understanding the nature of dark energy and its role in the expansion of the Universe.
\subsection{Methodology}

Markov Chain Monte Carlo (MCMC) methods have become indispensable tools in cosmology for exploring parameter spaces and estimating uncertainties. In the field of cosmology, MCMC techniques offer a powerful approach to sampling from high-dimensional posterior distributions, enabling us to infer cosmological parameters and understand the underlying models. MCMC algorithms have proven particularly valuable for fitting theoretical models to observed data, such as cosmic microwave background measurements, large-scale structure surveys, and type Ia supernova data. By comparing model predictions to observational data, MCMC methods allow us to constrain parameter values and quantify their uncertainties. The MCMC process involves constructing a Markov chain, where each point represents a specific set of parameter values. These values are randomly generated based on a proposal distribution and accepted or rejected based on the likelihood of the data given the model and prior information. Through iterations, the Markov chain converges to the target posterior distribution, providing samples that represent parameter estimates. The strength of MCMC lies in its ability to efficiently explore complex parameter spaces, even in the presence of multimodal or degenerate likelihood surfaces. By generating a large number of samples from the posterior distribution, we can estimate parameter uncertainties, derive credible intervals, and assess model goodness of fit. In cosmology, MCMC techniques have been extensively employed to investigate various topics, including the properties of dark matter and dark energy, inflationary models, primordial power spectra, and cosmological parameters like the Hubble constant, matter density, and dark energy equation of state.

\subsection{Data Discription}
\subsubsection{$H(z)$ measurements}
In our analysis, we have utilized the Hubble expansion rate data to impose more stringent constraints on our Dark Energy (DE) models. The Hubble expansion rate, denoted by $H(z)$, describes how the Universe's expansion rate changes with cosmic time or redshift, and it is a critical parameter for understanding the evolution of the cosmos. We can derive $H(z)$ data through two main methods: the clustering of galaxies and quasars by measuring the Baryon Acoustic Oscillations (BAO) in the radial direction \cite{gaztanaga2009clustering}, or the differential age method, which expresses the Hubble parameter as follows:

\begin{equation}
   H(z) = -\frac{1}{1+z} \frac{d z}{d t},
\end{equation}

where $\frac{d z}{d t}$ can be inferred from the redshift difference $\Delta z$ and the age difference $\Delta t$ between two passively evolving galaxies \cite{jimenez2002constraining}. For the current analysis, we have employed a compilation of 36 data points of the Hubble parameter, which are presented in table II of the reference \cite{bouali2019cosmological}. Each data point in the compilation is associated with its corresponding reference. To quantify the agreement between the model predictions and the observed values of $H(z)$, we utilize the $\chi_{H(z)}^{2}$ statistic, given by:

\begin{equation}
\chi_{H(z)}^{2} = \sum_{i=1}^{36} \frac{\left[H_{\text{th}}\left(z_{i}\right) - H_{\text{obs}}\left(z_{i}\right)\right]^{2}}{\sigma_{H\left(z_{i}\right)}^{2}},
\end{equation}

where $H_{\text{th}}$ represents the model predictions, $H_{\text{obs}}$ represents the measured values, and $\sigma_{H\left(z_{i}\right)}$ is the standard deviation associated with each data point at redshift $z_{i}$.

\subsubsection{The type Ia supernova dataset}
The study of Type Ia supernovae (SNIa) has played a crucial role in our understanding of cosmic accelerated expansion. SNIa has proven to be one of the most efficient methods for investigating the nature of dark energy, which remains one of the most intriguing and mysterious phenomena in cosmology. Over the past few years, a wealth of supernova data has been collected, providing valuable insights into the expansion of the Universe \cite{Pan1,Pan2,Pan3,Pan4,Pan5,pantheon+}. The recent release of a comprehensive dataset comprising 1701 SNIa observations is a significant milestone in this field. These observations span an extensive redshift range from  $0.001<z< 2.3$., allowing researchers to explore the Universe's expansion history over a broad cosmic time scale. SNIa are exceptionally luminous astronomical events, and they are considered to be nearly perfect standard candles. This unique property makes them valuable tools for measuring relative distances using the distance modulus. The distance modulus relates the intrinsic brightness of the supernova to its observed apparent brightness, enabling scientists to calculate precise distances to distant objects and thus probe the expansion rate of the Universe at different epochs. The chi-square values of the SNIa dataset are given by
\begin{equation}\label{eq_chi2}
 {\chi}_{{Pantheon+}}^2={\vec{D}}^{T}\hspace{0.1cm}.\hspace{0.1cm}{\bf C}_{{Pantheon+}}^{-1}\hspace{0.1cm}.\hspace{0.1cm}{\vec{D}},
\end{equation}
where ${\bf C}_{{Pantheon+}}$ is the covariance matrix provided with the Pantheon+ data, including both statistical and systematic uncertainties. and
\be
\vec{D}=m_{B i}-M-\mu_{\text{model}},
\ee
where $m_{B i}$ and $M$ denote the apparent, and the absolute magnitudes of SNIa, respectively. Furthermore, $\mu_{\text{model}}$ denotes the corresponding distance modulus as predicted by an assumed cosmological model, and defined as
\begin{equation}
\mu_{\text{model}}\left(z_{i}\right)=5\log_{10} \frac{D_L(z_{i})}{(H_0/c) Mpc} +25,
\end{equation}
where $H_0$ is the current value of the Hubble rate, and $D_L$ is the luminosity distance,  expressed, for a flat homogeneous and isotropic FLRW Universe,  as follows
\begin{equation}
D_L(z)=(1+z)H_0\int_{0}^{z}\frac{dz^{\prime}}{H\left(z^{\prime}\right)}.
\end{equation}

Unlike the Pantheon sample,  the degeneracy between the absolute magnitude $M$ and $H_0$ is broken in the Pantheon + dataset, due to the rewriting of the vector $\vec{D}$, appearing in  Eq. (\ref{eq_chi2}), in terms of the distance moduli of SNIa in the Cepheid hosts,  which allows to constrain $M$ independently via the relations
\begin{equation}
\vec{D}_{i}^{\prime}= \begin{cases}m_{B i}-M-\mu_{i}^{\text {Ceph }} & i \in \text { Cepheid hosts } \\ m_{B i}-M-\mu_{\text {model }}\left(z_{i}\right) & \text { otherwise },\end{cases}
\end{equation}
where $\mu_{i}^{\rm Ceph}$ stands for the distance modulus corresponding to the Cepheid host of the $i^{\text {th }}$ SNIa,  measured independently  with Cepheid calibrators. Therefore, Eq. (\ref{eq_chi2}) can be  rewritten as

\begin{equation}\label{eq_chi2_rew}
\chi_{SN}^2 = \vec{D'}^T \cdot {\bf C}_{\text{Pantheon+}}^{-1} \cdot \vec{D'}
\end{equation}

We also have involve 162 Gamma Ray Bursts (GRBs) ranging in the redshift from $1.44$ to $8.1$ \cite{GRB}. GRBs are incredibly bright and short-lived events, making them observable even from the farthest reaches of the visible Universe. In addition to the GRBs, We also included 24 observations of compact radio quasars \cite{quasers}. These quasars were found in the redshift range of $0.46$ to $2.76$ and were identified through a survey of 613 ultra-compact radio sources at GHz frequencies using Very Long Baseline Interferometry (VLBI). Both GRBs and compact radio quasars are essential objects in cosmology. GRBs, being among the most luminous events in the Universe, provide valuable information about the distant regions of the cosmos. On the other hand, compact radio quasars offer insights into the properties of active galactic nuclei and their evolution over cosmic time. The combination of these datasets, along with other cosmological probes like Type Ia supernovae, allows us to gain a deeper understanding of the expansion history and the nature of dark energy.

\subsubsection{Baryon Acoustic Oscillations (BAO)}
We have used 17 BAO data points  (please see table 1 of this work \cite{benisty2021testing}) measures from \cite{benisty2021testing}) measures from \cite{baonew1,baonew2,bao3,bao4,bao5,bao6,bao7,bao8,bao9,bao10,bao11,bao12} a whole catalog of BAO (333) measurements have been selected because considering the whole catalog of BAO might result in significant inaccuracy because of data correlations; hence, a representative subset to minimize errors. Transverse BAO experiments produce measurements. of $D_H(z) / r_d=c / H(z)r_d$ with a comoving angular diameter distance.\cite{bao13,bao14}.

\begin{equation}
D_M=\frac{c}{H_0} S_k\left(\int_0^z \frac{d
z^{\prime}}{E\left(z^{\prime}\right)}\right) \text {, }
\end{equation}

where

\begin{equation}
S_k(x)= \begin{cases}\frac{1}{\sqrt{\Omega_k}} \sinh
\left(\sqrt{\Omega_k} x\right) & \text { if } \quad \Omega_k>0 \\
x & \text { if } \quad \Omega_k=0 \\ \frac{1}{\sqrt{-\Omega_k}}
\sin \left(\sqrt{-\Omega_k} x\right) & \text { if } \quad
\Omega_k<0 .\end{cases}
\end{equation}

We also consider the angular diameter distance $D_A=$ $D_M /(1+z)$ and the $D_V(z)/r_d$. This corresponds to the combination of the BAO peaked coordinates and the sound horizon $r_d$ at the drag epoch. Furthermore, we could immediately derive "line-of-sight" (or "radial") observations from the Hubble parameter.

\begin{equation}
D_V(z) \equiv\left[z D_H(z) D_M^2(z)\right]^{1/3} \text {. }
\end{equation}

    \begin{table}
        \begin{center}
            \begin{tabular}{|c|c|c|c|c|c|c|c|}
                \hline
                \multicolumn{8}{|c|}{MCMC Results} \\
                \hline
                Model & Parameters & Priors & Bestfit Value & Model & Parameters & Priors & Bestfit Value \\
                \hline
               $  \Lambda$CDM &$H_{0}$&[50.,100.] &$69.854848_{-1.259100}^{+1.259100}$ & $  \Lambda$CDM & $H_0 $ &[50.,100.]&$69.158620_{-0.454345}^{+0.354533}$ \\
                &  $\Omega_{\mathrm{m0}}$ & $[0.,1.]$ &$0.268654_{-0.015435}^{+0.012822}$ & & $\Omega_{\mathrm{m0}}$ &$[50.,100.]$&$0.268654_{-0.012822}^{+0.035225}$ \\
                \hline
                MCJG & $H_0 $ & [50.,100.] &$69.365373_{-1.493749}^{+2.493749}$ & MCAG & $H_0 $ &[50.,100.]&$69.158620_{-1.343566}^{+1.593749}$ \\

                & $\Omega_{\mathrm{m0}}$ &[0.,1.] &$0.203134_{-0.045126}^{+0.0356345}$& & $\Omega_{\mathrm{m0}}$ &[0.,1.] &$0.322633_{-0.0545433}^{+0.043225}$ \\

                &$\Omega_{\mathrm{r0}}$ &[0.,1.] &$0.293748_{-0.001282}^{+0.0034524}$& & $\Omega_{\mathrm{r0}}$ &[0.,1.]& $0.489371_{-0.023543}^{+0.0443255}$ \\

                &$\Omega_{\mathrm{k0}}$ &[0.,1.] &$0.134637_{-0.001233}^{+0.004322}$& & $\Omega_{\mathrm{k0}}$ &[0.,1.]& $0.285599_{-0.0445635}^{+0.045126}$ \\

                &$\Omega_{\mathrm{CJ0}}$ & [0.,1.]&$0.477354_{-0.003242}^{+0.001787}$ & & $\Omega_{\mathrm{CAO}}$ & [0.,1.]&$0.700436_{-0.07655}^{+0.06754}$ \\

                & $A$ &[-10.,-9.]&$-9.234245 _{-0.0084643}^{+0.009433}$ & & $A$ &[-8.,-7.]& $-7.543525_{-0.045324}^{+0.044255}$  \\

                & $\alpha$ &[1.5,2.5]&$2.093362 _{-0.008374}^{+0.009647}$ & & $\alpha$ &[1.5,2.5]& $2.142141_{-0.044232}^{+0.065435}$ \\

                & $\beta$ &[-0.1,0.1]&$-0.025066_{-0.053542}^{+0.043245}$ & & $B_{s}$ &[2.,3.]& $2.534432_{-0.0155433}^{+0.069934}$ \\

                & $\gamma$ &[0.,1.]&$0.567318 _{-0.086122 }^{+0.095311 }$ & & $b$ & [1.,2.]&$1.481343_{-0.094873}^{+0.067483}$  \\

                & $A_{s}$ &[0.5,1.5]&$0.959363 _{-0.009647}^{+0.006647}$  & & $c$ & [0.,1.]&$0.694735_{-0.067843}^{+0.093853}$ \\

                & $K$ &[0.,1.]&$0.692843 _{-0.003532}^{+0.006432}$ & & $\beta$ & [0.,0.5]&$0.0377415_{-0.067785}^{+0.07584}$\\

                & & & & & $\gamma$ &[0.,1.] &$0.6427995_{-0.079584}^{+0.069383}$\\
                \hline
            \end{tabular}
        \end{center}
    \end{table}

\newpage
    \begin{figure}
        \centerline{\includegraphics[scale=0.35]{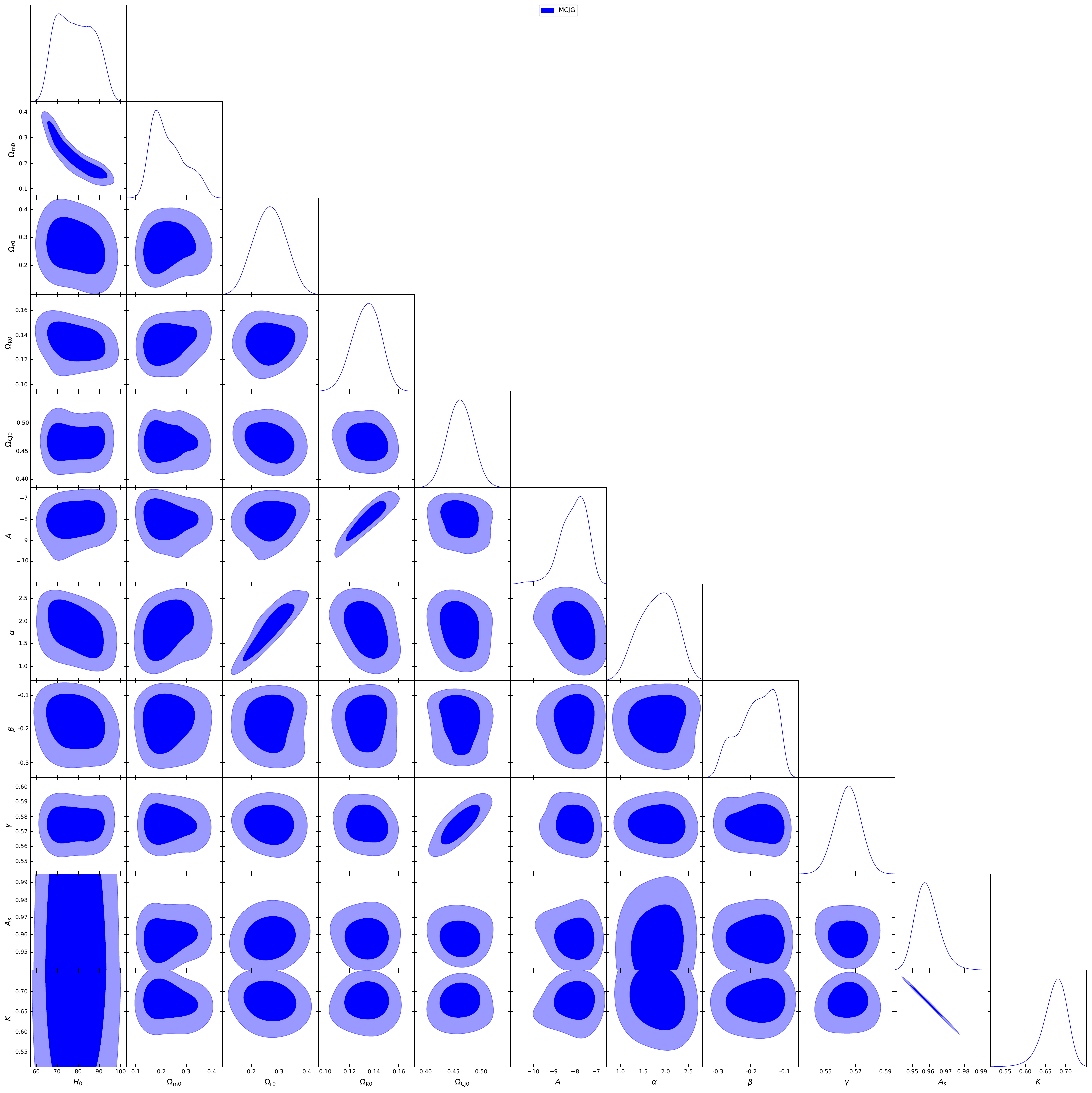}}
        \caption{The above figure shows the MCMC confidence contours at 1$\sigma$ and 2$\sigma$ of MCJG Model}
        \label{MCJG}
    \end{figure}

    \begin{figure}
        \centerline{\includegraphics[scale=0.32]{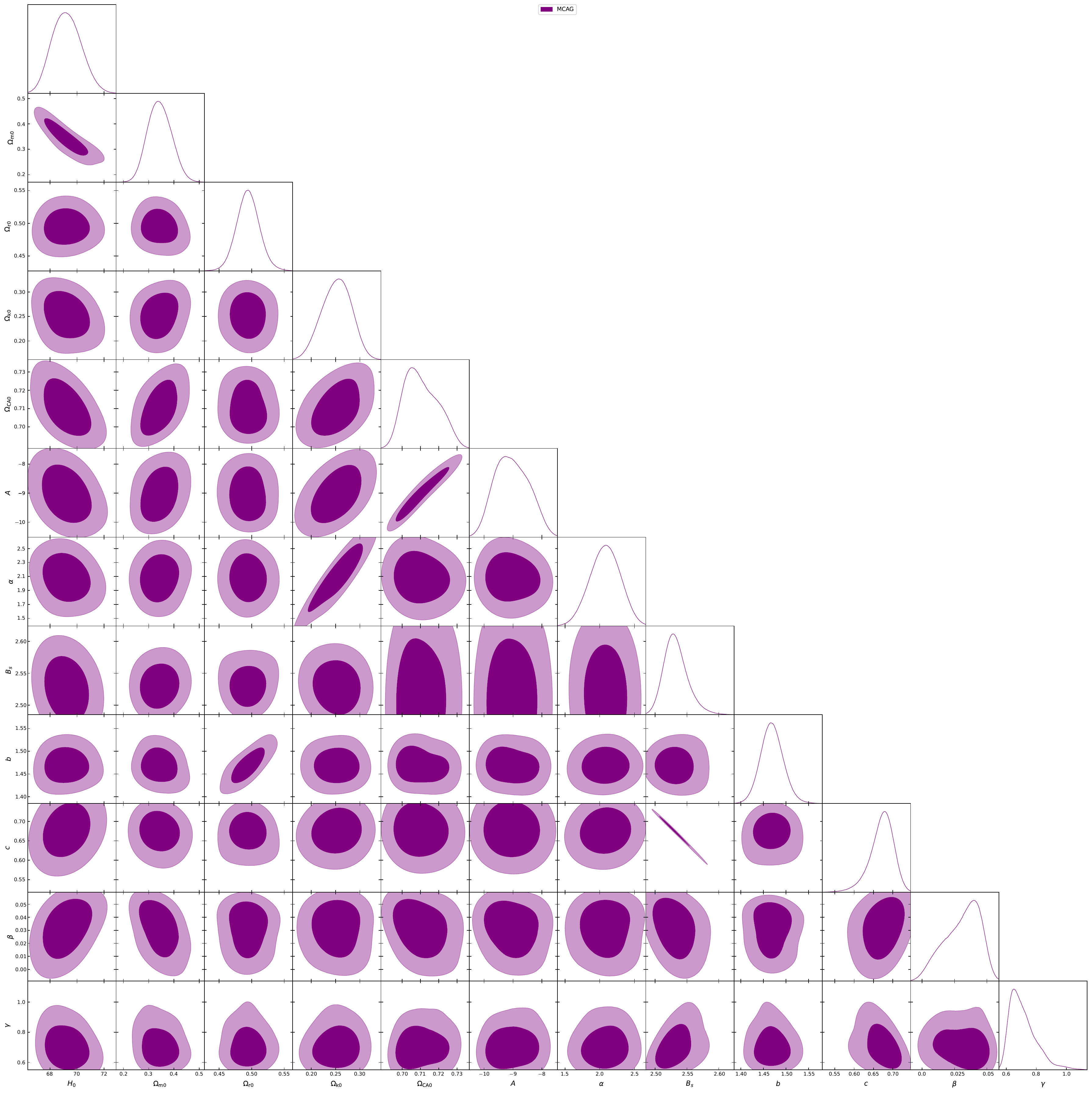}}
        \caption{The above figure shows the MCMC confidence contours at 1$\sigma$ and 2$\sigma$ of MCAG Model}
        \label{MCAG}
    \end{figure}

\clearpage
\section{Observational and theoretical comparisons of the Hubble Function and Distance Modulus Function}\label{secV}
After obtaining the best-fit values for the model parameters of the MCJG and MCAG Models, it is crucial to compare these models with the widely accepted $\Lambda$CDM model, which has demonstrated remarkable consistency with various observational datasets. The $\Lambda$CDM model is considered a robust framework for describing the evolution of the Universe and serves as a reference point for assessing the viability and reliability of alternative models. By comparing MCJG and MCAG Models with the $\Lambda$CDM model, we can investigate the deviations and discrepancies between the two and gain insights into the unique features and behaviors of MCJG and MCAG Models. This comparative analysis allows us to examine how MCJG and MCAG Models differ from the $\Lambda$CDM model in terms of the expansion rate, matter content, and overall dynamics of the Universe. Analyzing the deviations between MCJG and MCAG Models and the $\Lambda$CDM model provides valuable information about the strengths and limitations of MCJG and MCAG Models. It helps us understand the specific aspects of MCJG and MCAG Models that introduce variations compared to the well-established $\Lambda$CDM framework. By identifying these deviations, we can gain insights into the potential implications of MCJG and MCAG Models for our understanding of the Universe and explore novel cosmological dynamics and characteristics. Furthermore, the comparison with the $\Lambda$CDM model allows us to assess the goodness-of-fit of MCJG and MCAG Models to observational data. By evaluating the level of agreement between MCJG and MCAG Models models and the $\Lambda$CDM framework, we can determine the validity and reliability of MCJG and MCAG Models in light of the available observational constraints.

\subsection{Comparison of MCJG and MCAG with Hubble Measurements}
\begin{figure}[!htb]
   \begin{minipage}{0.49\textwidth}
     \centering
   \includegraphics[scale=0.4]{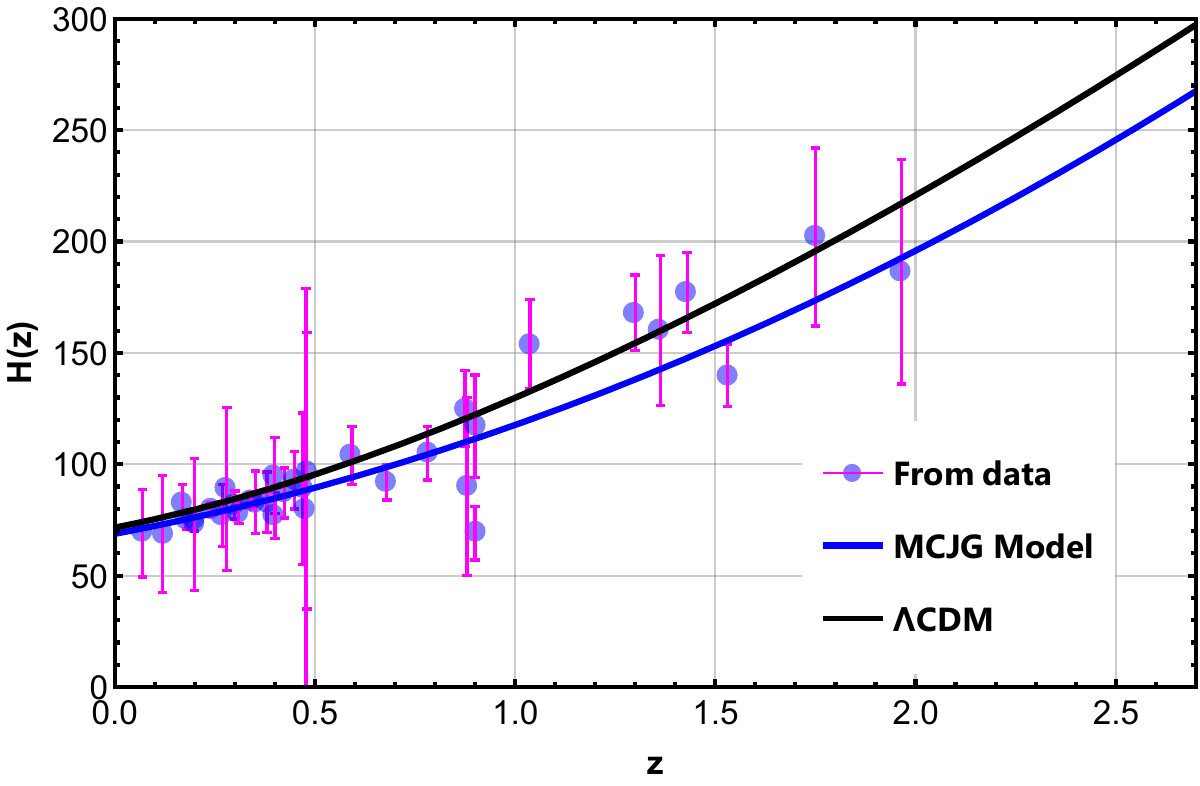}
\caption{The figure shows that the theoretical curve of the Hubble function $H(z)$ of MCJG Model shown in blue line and $\Lambda$CDM model shown in black line with $\Omega_{\mathrm{m0}}=$ 0.3 and $\Omega_\Lambda =$ 0.7 against the 36 Hubble measurements are shown in blue dots wit their corresponding error bars in purple line. }\label{H(z) Model 1}
   \end{minipage}\hfill
   \begin{minipage}{0.49\textwidth}
     \centering
    \includegraphics[scale=0.4]{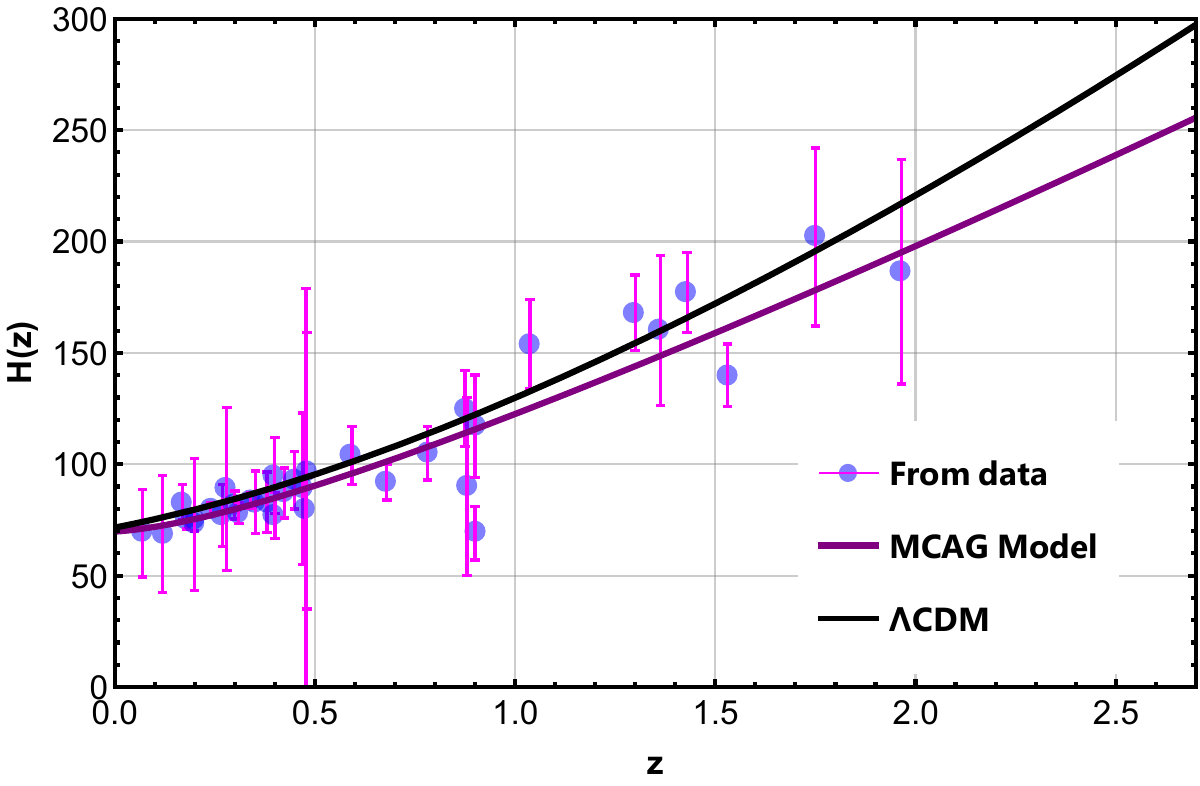}
\caption{The figure shows that the theoretical curve of the Hubble function $H(z)$ of MCAG Model shown in purple line and $\Lambda$CDM model shown in black line with $\Omega_{\mathrm{m0}}=$ 0.3 and $\Omega_\Lambda =$ 0.7, against 36 Hubble measurements are shown in blue dots wit their corresponding error bars in purple line.}\label{H(z) Model 2}
   \end{minipage}
\end{figure}

To evaluate the agreement between MCJG and MCAG models with observational data, we compared their predictions to the Hubble data, alongside the well-established $\Lambda$CDM model. The results are presented in Fig:- \ref{H(z) Model 1} and Fig:- \ref{H(z) Model 2}. The figures clearly show that both MCJG and MCAG models fit the Hubble data quite well. The data points obtained from observations match very well with the predictions of our models, indicating that both models accurately explain how the Universe is expanding. The fact that both MCJG and MCAG models match the Hubble data provides strong support for their viability and indicates that they capture essential aspects of cosmic expansion. These findings demonstrate the satisfactory agreement of both models with the Hubble data, underscoring their potential to provide meaningful insights into the dynamics and evolution of the Universe within the framework of our proposed models. Such a positive outcome reinforces the credibility and significance of our research and its ability to contribute to a better understanding of the Universe's evolution.

\subsection{Comparison of MCJG and MCAG with type Ia supernova}
\begin{figure}[!htb]
   \begin{minipage}{0.49\textwidth}
     \centering
   \includegraphics[scale=0.4]{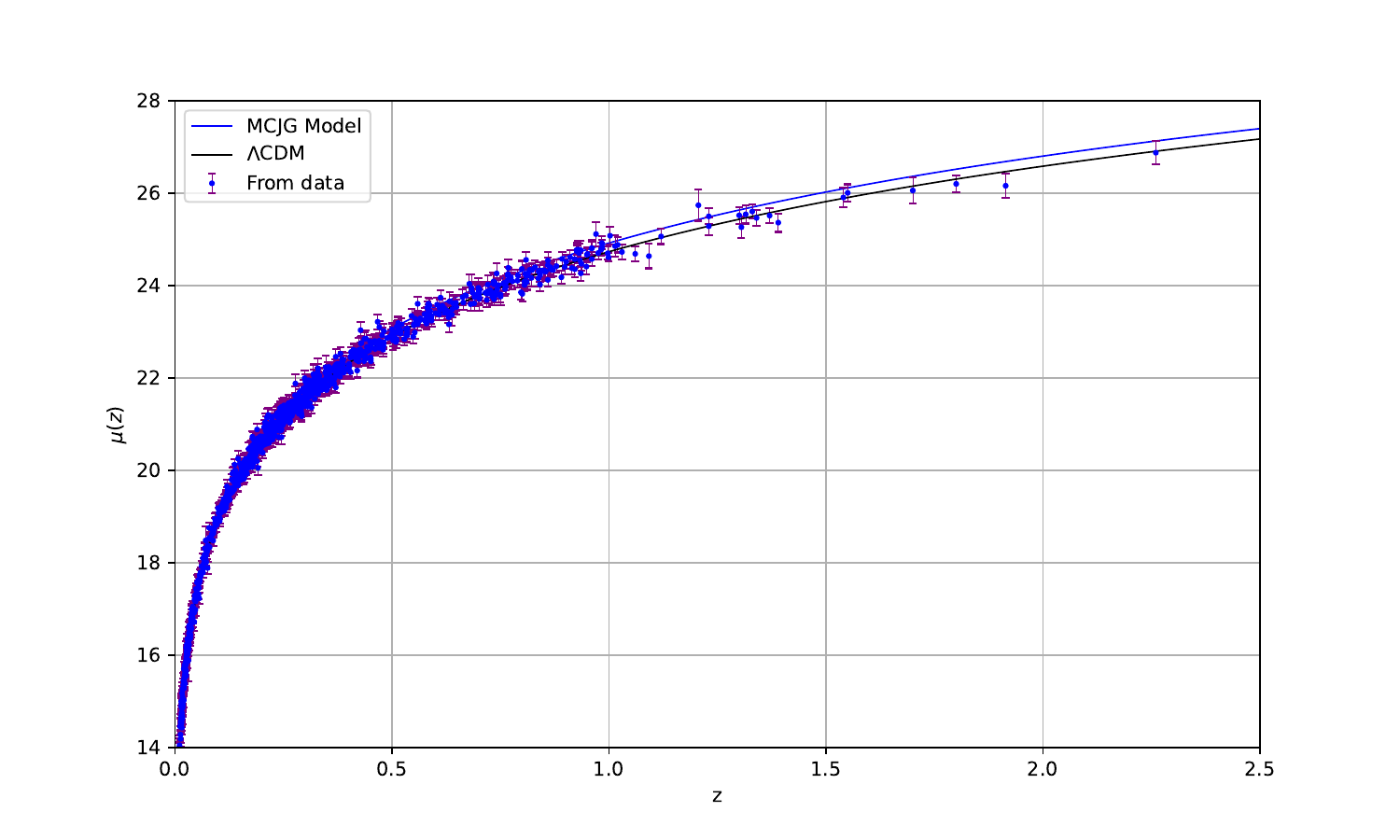}
\caption{Theoretical curve of distance modulus $\protect\mu(z) $ of the MCJG is shown in blue line and the $\Lambda$CDM model is shown in the black line with $\Omega_{\mathrm{m0}}=$ 0.3 and $\Omega_\Lambda =$ 0.7, against type Ia supernova data are shown in blue dots with their corresponding errors bars shown in purple.}\label{mu(z) Model 1}
   \end{minipage}\hfill
   \begin{minipage}{0.49\textwidth}
     \centering
    \includegraphics[scale=0.4]{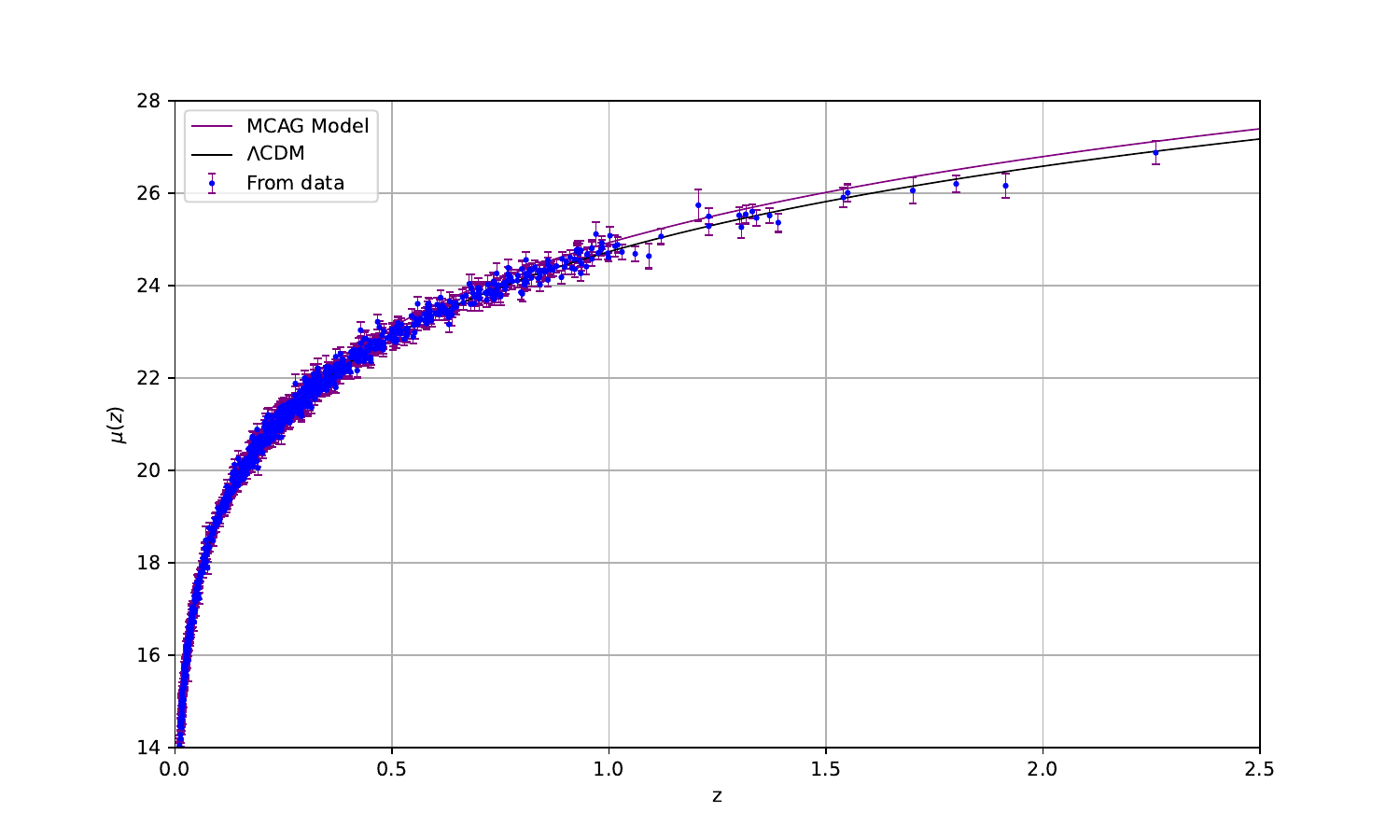}
\caption{Theoretical curve of distance modulus $\protect\mu(z) $ of the MCAG is shown in purple line and the $\Lambda$CDM model is shown in the black line with $\Omega_{\mathrm{m0}}=$ 0.3 and $\Omega_\Lambda =$ 0.7, against type Ia supernova data are shown in blue dots with their corresponding errors bars shown in purple.}
\label{mu(z) Model 2}
   \end{minipage}
\end{figure}

In this analysis, we have compared two distance modulus functions, labeled as MCJG and MCAG, with the type Ia supernova dataset, which consists of 1701 data points, as well as the standard $\Lambda$CDM model. The comparison results are illustrated in Fig:- \ref{mu(z) Model 1} and Fig:- \ref{mu(z) Model 2}. These figures shows that both MCJG and MCAG provide an excellent fit to the type Ia supernova dataset and the $\Lambda$CDM model. This means that our models closely match the observed distance measurements, and they are consistent with the data we have collected. This comparison with real observational data provides strong support for the validity and trustworthiness of our models in explaining the phenomena we observe in the Universe. It gives us confidence that both MCJG and MCAG accurately represent the expansion history of the Universe, as indicated by the type Ia supernova dataset.

\subsection{Relative difference between MCJG and MCAG with $\Lambda$CDM}
\begin{figure}[!htb]
   \begin{minipage}{0.49\textwidth}
     \centering
   \includegraphics[scale=0.43]{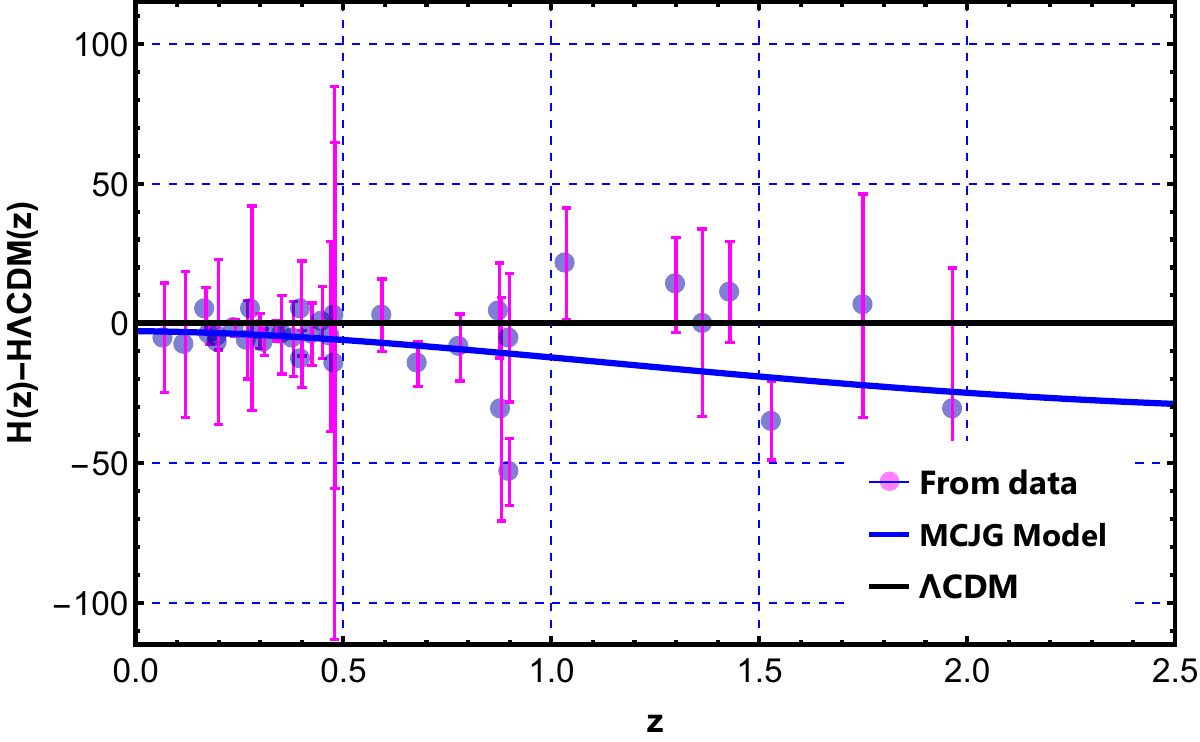}
\caption{The variation of the difference between MCJG shown in the blue line, and the $\Lambda$CDM model shown in the black line with $\Omega_{\mathrm{m0}}=$ 0.3 and $\Omega_\Lambda =$ 0.7, as a function of the
redshift $z$ against the 36 Hubble measurements are shown in blue dots with their corresponding error bars shown in purple line.}\label{h(z)diff1}
   \end{minipage}\hfill
   \begin{minipage}{0.49\textwidth}
     \centering
    \includegraphics[scale=0.43]{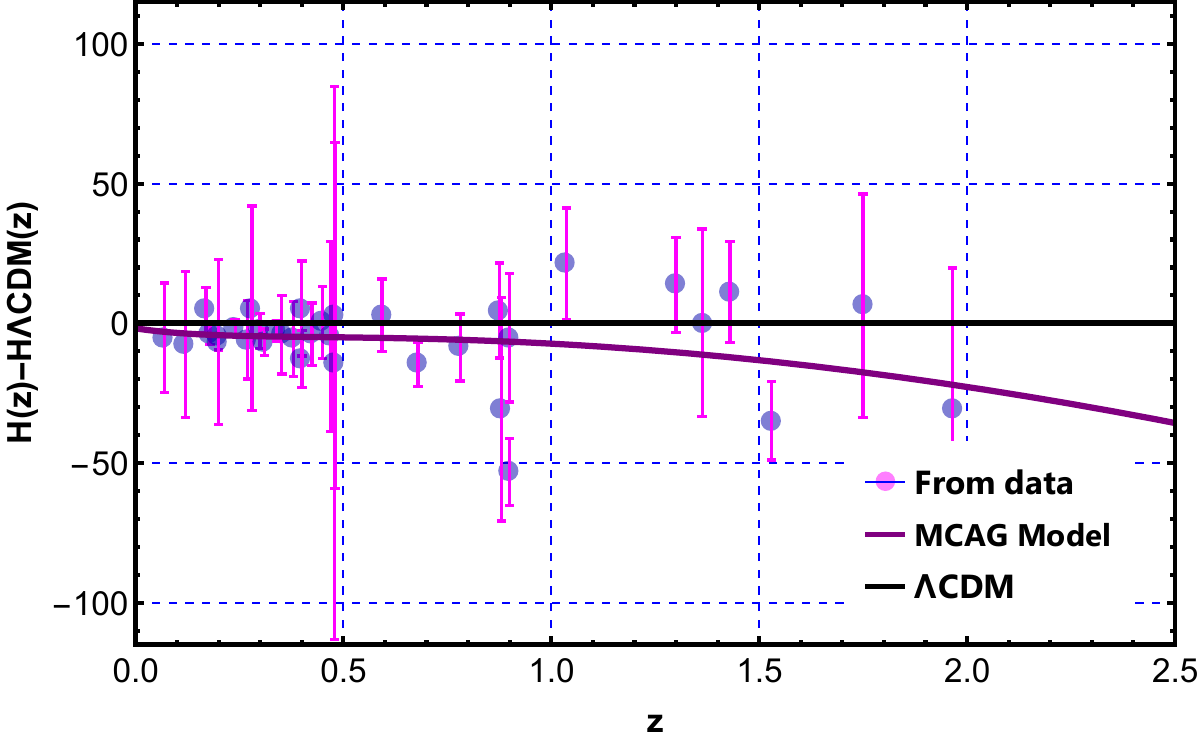}
\caption{The variation of the difference between MCAG shown in the purple line, and the $\Lambda$CDM model shown in the black line with $\Omega_{\mathrm{m0}}=$ 0.3 and $\Omega_\Lambda =$ 0.7, as a function of the
redshift $z$ against the 36 Hubble measurements are shown in blue dots with their corresponding error bars shown in purple line.}\label{h(z)diff2}
   \end{minipage}
\end{figure}

We have examined the relative difference between MCAJG and MCAG, and the standard $\Lambda$CDM paradigm. These comparisons are visualized in Fig:- \ref{h(z)diff1} and Fig:- \ref{h(z)diff2}. The figures reveal interesting insights regarding the performance of the typical $\Lambda$CDM model and the two alternative models for different redshifts. Specifically, for redshifts $z$ less than 1, all three models demonstrate very similar behavior, indicating a close agreement in their predictions. However, as we move to higher redshifts ($z > 1$), discrepancies begin to emerge between the two alternative models and the $\Lambda$CDM model. The differences become more pronounced as the redshift increases. This suggests that at higher redshifts, the alternative models deviate from the predictions of the $\Lambda$CDM model.

\newpage
\section{Geometrical dynamics of the models}\label{secVI}

\subsection{Cosmographic analysis}
Cosmography parameters \cite{visser2004jerk,visser2005cosmography,capozziello2011cosmography} provide a valuable framework for analyzing and comparing theoretical cosmological models with observational data. They offer a comprehensive set of quantities that describe the expansion history of the Universe and can be used to test the consistency of different cosmological models. The cosmography approach is based on a Taylor expansion of the scale factor, which represents the size of the Universe, around a reference redshift or time. By expanding the scale factor up to a certain order, we can obtain a series of cosmography parameters that capture important aspects of the cosmic evolution. These parameters are derived from observables such as the Hubble parameter and its derivatives. Some of the commonly used cosmography parameters include the Hubble parameter ($H(z)$), deceleration parameter ($q(z)$), jerk parameter ($j(z)$), and snap parameter ($s(z)$), among others. These parameters can be derived from the expansion history of the Universe and provide a quantitative description of its behavior. They allow us to extract meaningful information about the underlying cosmological model and compare it with observational data.

\subsubsection{deceleration parameter}
The deceleration parameter is a fundamental cosmological quantity that describes the rate at which the expansion of the Universe is slowing down or speeding up. It provides valuable insights into the dynamics of the cosmic expansion and plays a crucial role in understanding the past and future behavior of the Universe. In the context of Friedmann-Robertson-Walker cosmology, the deceleration parameter, denoted as $q(z)$, is defined as the negative ratio of the second derivative of the scale factor $a(t)$ to its first derivative, normalized by the square of the Hubble parameter $H(z)$:

\begin{equation}
q(z) = -\frac{a''(t)}{a(t)H(t)^2} 
\end{equation}

Here, $a(t)$ represents the scale factor of the Universe, which describes how the distances between galaxies change with time, and $H(t)$ is the Hubble parameter, which characterizes the expansion rate of the Universe at a given time. The deceleration parameter $q(z)$ determines the behavior of the cosmic expansion. If $q(z) < 0$, the Universe is experiencing accelerated expansion, indicating that the expansion rate is increasing over time. Conversely, if $q(z) > 0$, the Universe is undergoing decelerated expansion, with the expansion rate decreasing over time. Studying the deceleration parameter is essential for understanding the evolution of the Universe and the influence of different components such as matter, radiation, or dark energy on its expansion dynamics. References such as \cite{b1,b2,b3,b4,b5} have employed the deceleration parameter to analyze and explain the evolutionary phases of the Universe in different cosmological models.

\begin{figure}[!htb]
   \begin{minipage}{0.49\textwidth}
     \centering
   \includegraphics[scale=0.43]{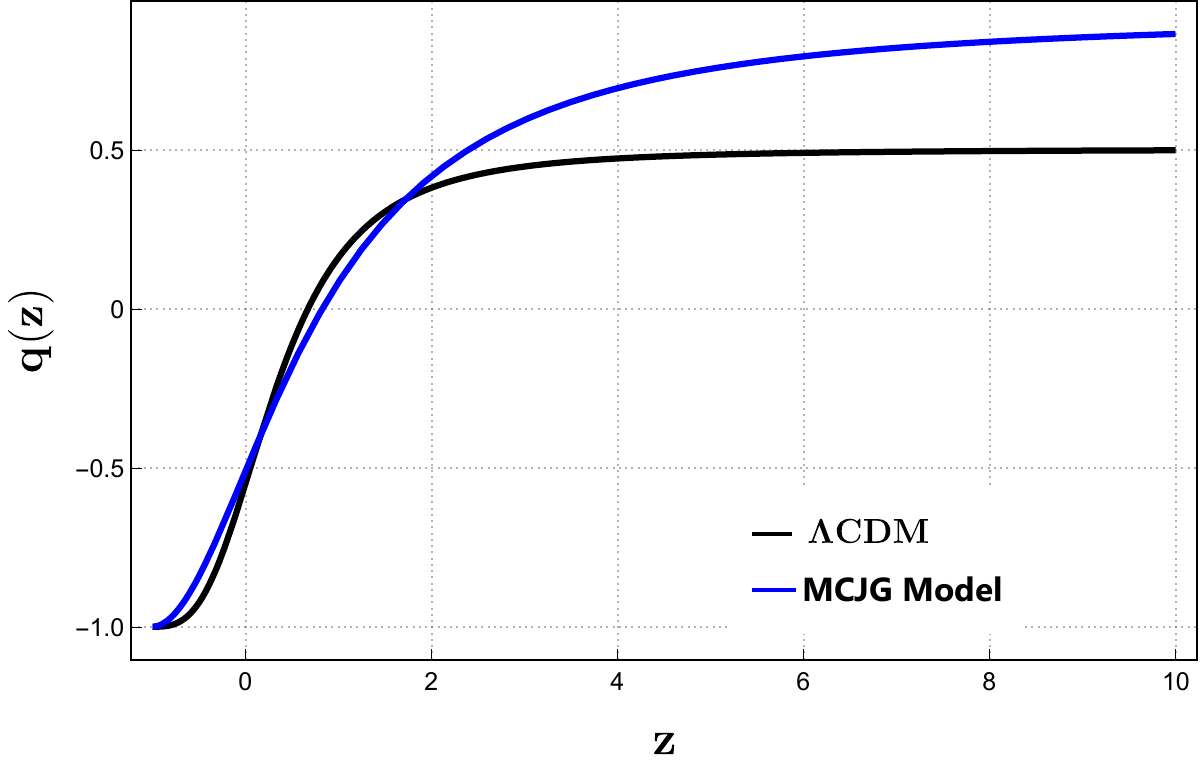}
\caption{Evolution of deceleration parameter with respect to the redshift of MCJG Model.}\label{q(z) Model 1}
   \end{minipage}\hfill
   \begin{minipage}{0.49\textwidth}
     \centering
    \includegraphics[scale=0.43]{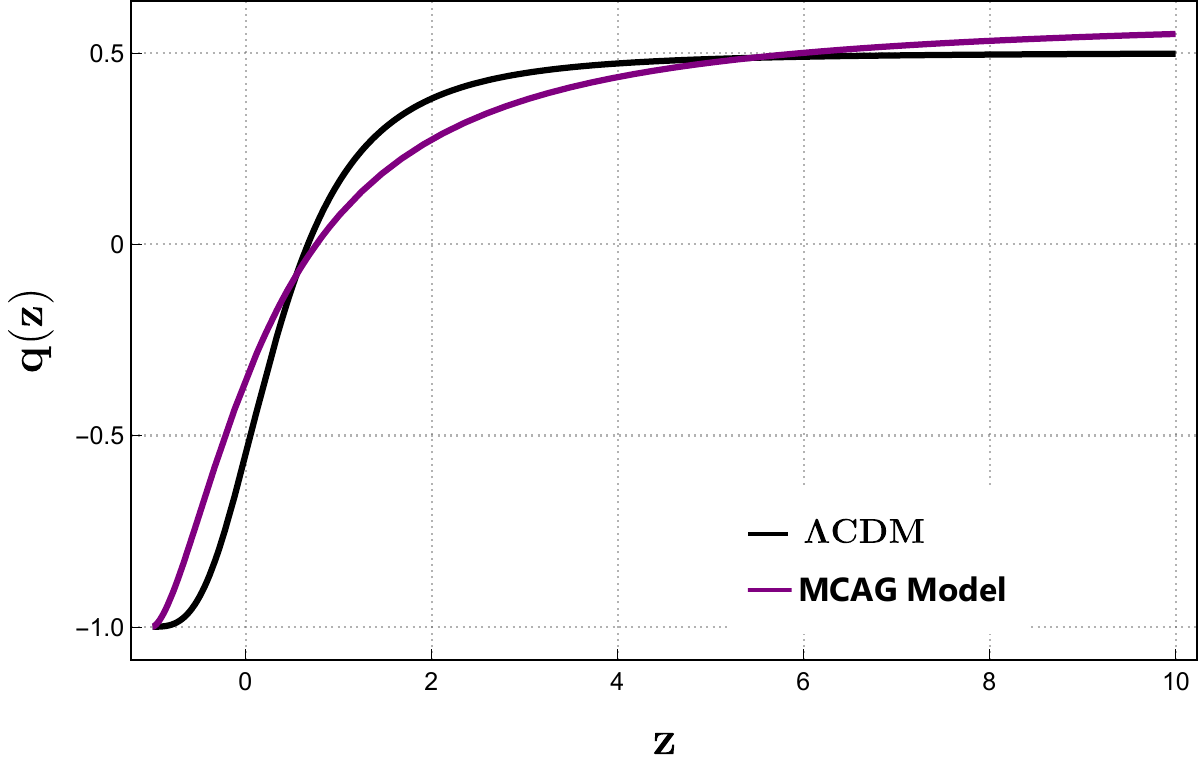}
\caption{Evolution of deceleration parameter with respect to the redshift of MCAG Model.}\label{q(z) Model 2}
   \end{minipage}
\end{figure}

\subsubsection{Jerk parameter}
The concept of jerk parameters in cosmology provides valuable insights into the dynamics of the Universe and the nature of dark energy. Jerk parameters offer novel diagnostics for dark energy, allowing us to study the higher derivatives of the scale factor and explore deviations from the standard cosmological model. In cosmology, the dimensionless jerk parameter, sometimes referred to as jolt, extends the standard cosmological parameters, such as the scale factor $a(t)$ and the deceleration parameter $q$. It arises from the fourth term of a Taylor series expansion of the scale factor around a reference time $t_0$. Mathematically, it can be expressed as:

\begin{equation}
\frac{a(t)}{a_0} = 1 + H_0(t - t_0) - \frac{1}{2}q_0H_0^2(t - t_0)^2 + \frac{1}{6}j_0H_0^3(t - t_0)^3 + O\left[(t - t_0)^4\right],
\end{equation}

where the subscript $0$ denotes the present values. The jerk parameter $j$ characterizes the third-order derivative of the expansion factor with respect to cosmic time and offers an absolute measure to explore departures from the standard $\Lambda$CDM model. It can be expressed as:

\begin{equation}
j = \frac{1}{a}\frac{d^3a}{d\tau^3}\left[\frac{1}{a}\frac{da}{d\tau}\right]^{-3} = q(2q + 1) + (1 + z)\frac{dq}{dz},
\end{equation}

where $\tau$ represents the cosmic time and $z$ is the redshift. This expression provides a direct way to determine the jerk parameter by relating it to the deceleration parameter $q$ and its derivative with respect to redshift. The jerk parameter is a valuable diagnostic tool in cosmology as it helps determine the most suitable candidate for the physical interpretation of cosmic dynamics, particularly in relation to dark energy models. Different values of the jerk parameter offer insights into the correlation between various dark energy proposals and standard Universe models. In the flat $\Lambda$CDM model, the jerk parameter takes the value $j = 1$. Deviations from this value indicate departures from the standard model and provide a way to explore transitions between different phases of cosmic acceleration.

\begin{figure}[!htb]
   \begin{minipage}{0.49\textwidth}
     \centering
   \includegraphics[scale=0.42]{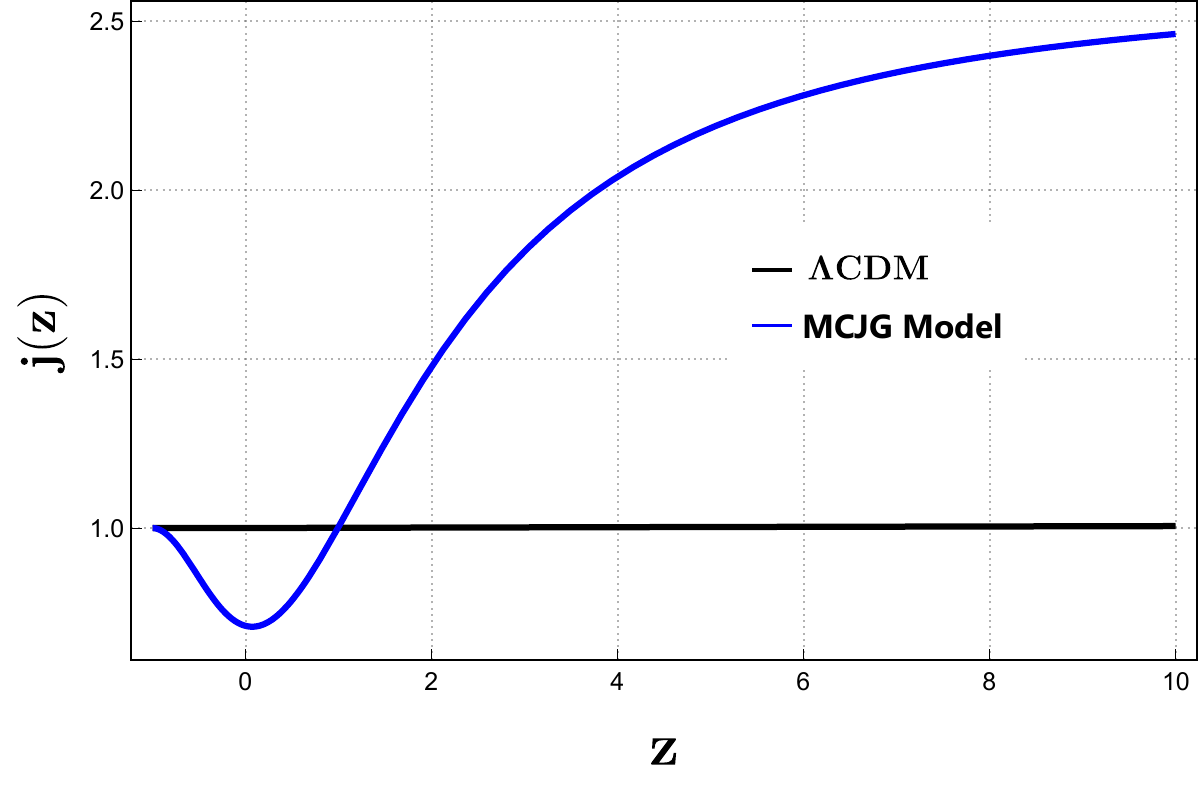}
\caption{Evolution of jerk parameter with respect to the redshift of MCJG Model.}\label{j(z) Model 1}
   \end{minipage}\hfill
   \begin{minipage}{0.49\textwidth}
     \centering
    \includegraphics[scale=0.42]{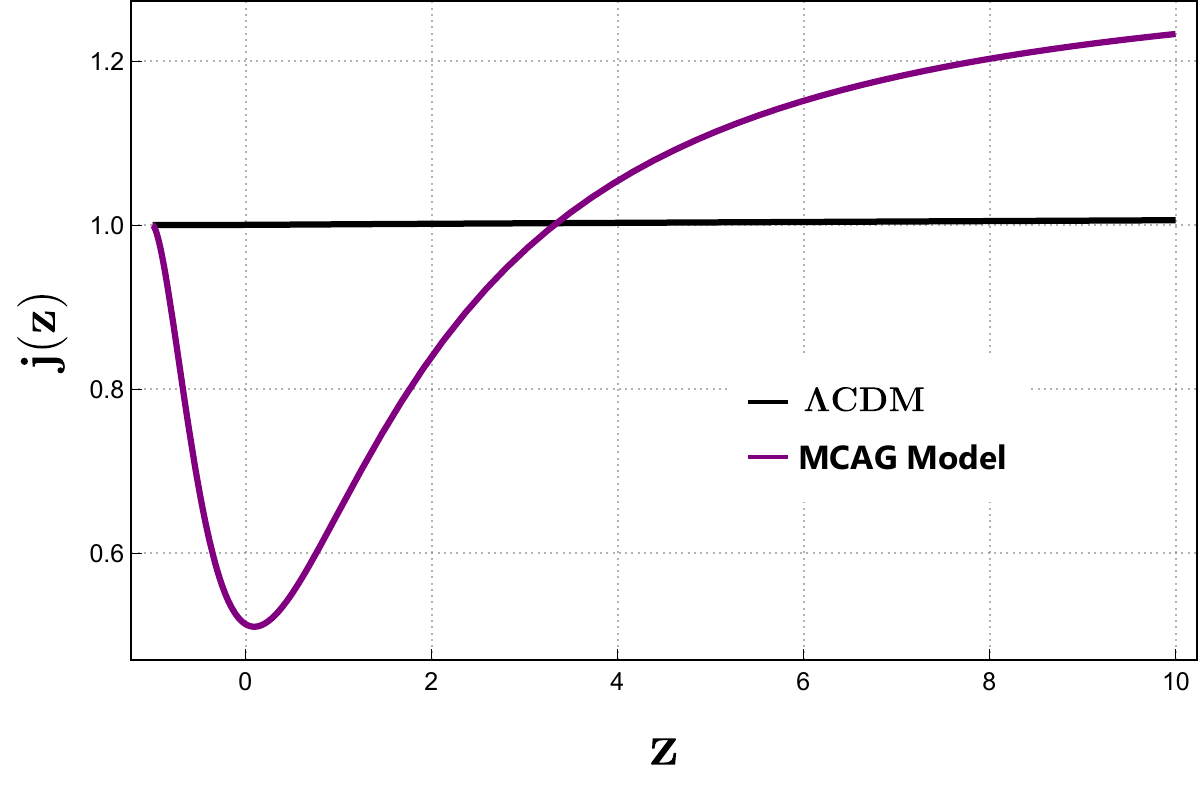}
\caption{Evolution of jerk parameter with respect to the redshift of MCAG Model.}\label{j(z) Model 2}
   \end{minipage}
\end{figure}

\subsubsection{snap parameter}
The snap parameter is a measure of the fifth time derivative of the expansion factor in cosmology. It is a dimensionless quantity that provides information about the curvature of the Universe and the dynamics of its expansion. The snap parameter is associated with the term in the Taylor expansion of the expansion factor beyond the third order. In the context of the Taylor expansion of the expansion factor, the snap parameter represents the fourth-order term. The expansion factor is expressed as a series expansion around a reference time $t_0$, with the snap parameter denoted as $s_0$. The expansion factor can be written as:

\begin{equation}
\frac{a(t)}{a_0} = 1 + H_0(t - t_0) - \frac{1}{2}q_0H_0^2(t - t_0)^2 + \frac{1}{6}j_0H_0^3(t - t_0)^3 + \frac{1}{24}s_0H_0^4(t - t_0)^4 + O\left[(t - t_0)^5\right]
\end{equation}

Here, $a(t)$ represents the expansion factor at time $t$, $a_0$ is the expansion factor at the reference time $t_0$, $H_0$ is the present-day Hubble constant, $q_0$ is the deceleration parameter, $j_0$ is the jounce parameter (related to the snap parameter), and $s_0$ is the snap parameter. The higher-order terms beyond the fourth order are neglected in this expansion. The snap parameter $s$ can also be expressed in terms of the deceleration parameter $q$ and the jounce parameter $j$ as:

\begin{equation}
s = \frac{1}{a}\frac{d^4a}{d\tau^4}\left[\frac{1}{a}\frac{da}{d\tau}\right]^{-4} = \frac{j - 1}{3\left(q - \frac{1}{2}\right)}
\end{equation}

For the flat $\Lambda$CDM model, where $j = 1$, the snap parameter simplifies to $s = -(2 + 3q)$. In this case, the deviation of the quantity $\frac{ds}{dq}$ from $-3$ quantifies how the cosmic evolution diverges from the dynamics predicted by the $\Lambda$CDM model.

\begin{figure}[!htb]
   \begin{minipage}{0.49\textwidth}
     \centering
   \includegraphics[scale=0.42]{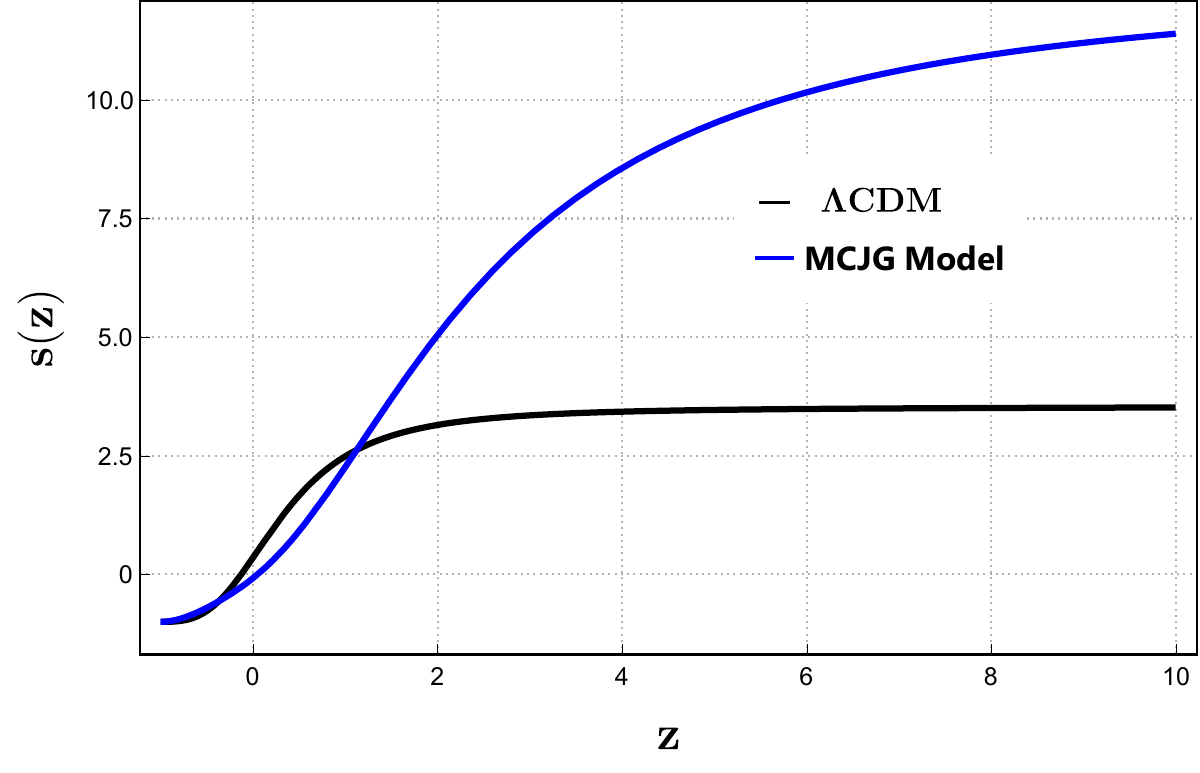}
\caption{Evolution of snap parameter with respect to the redshift of MCJG Model.}\label{s(z) Model 1}
   \end{minipage}\hfill
   \begin{minipage}{0.49\textwidth}
     \centering
    \includegraphics[scale=0.42]{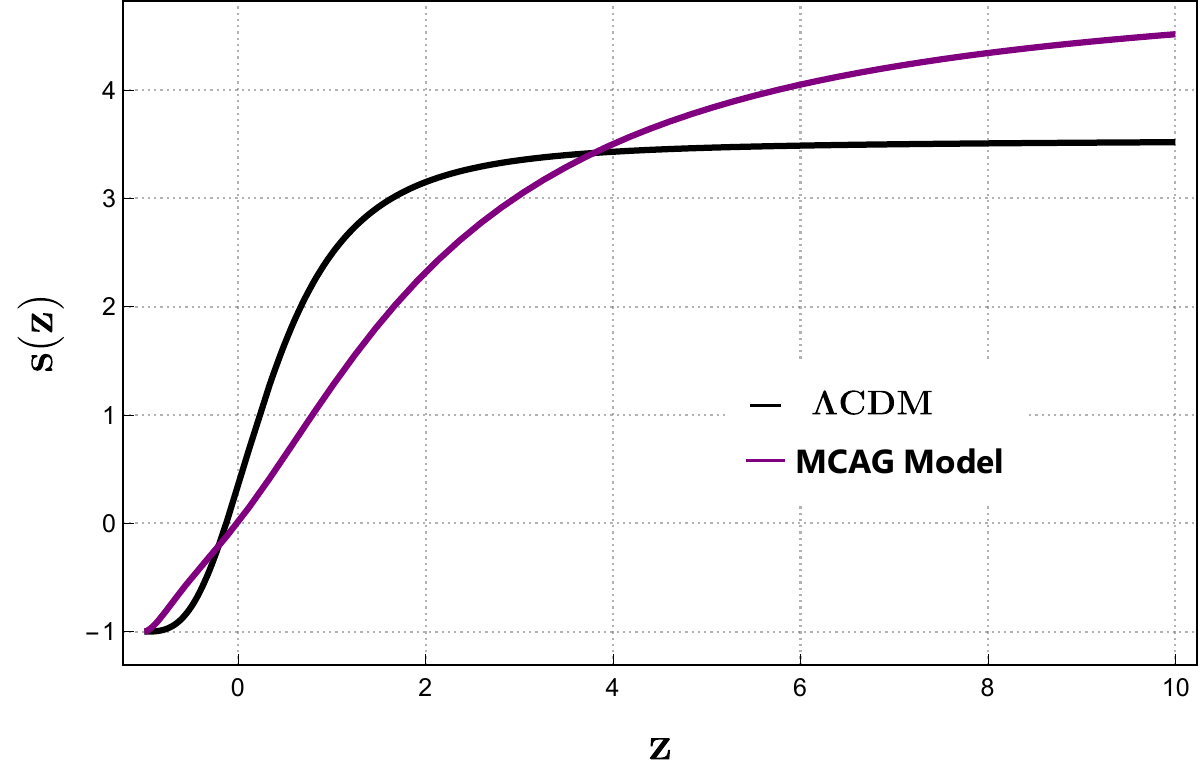}
\caption{Evolution of snap parameter with respect to the redshift of MCAG Model.}\label{s(z) Model 2}
   \end{minipage}
\end{figure}

\section{Statefinder Diagnostic}\label{secVII}
The statefinder diagnostic pair, consisting of the parameters $r$ and $s$, is a widely-used tool in cosmology for studying and understanding various dark energy (DE) models. It provides a dimensionless way to analyze the cosmic properties of DE, independent of specific models, by examining higher-order derivatives of the scale factor. This diagnostic pair was introduced in the literature ( \cite{state1,state2,state3,stste4} ). The parameter $r$ in the statefinder diagnostic represents the ratio of the third derivative of the scale factor to the cube of the Hubble parameter. It quantifies the acceleration dynamics of the Universe. On the other hand, the parameter $s$ is a linear combination of $r$ and the deceleration parameter $q$. It characterizes the cosmic evolution and the nature of DE. The statefinder diagnostic pair can be calculated using the following expressions:

\begin{equation}
r = \frac{\dddot{a}}{a H^3}, \quad s = \frac{r - 1}{3\left(q - \frac{1}{2}\right)},
\end{equation}

Here, $H$ is the Hubble parameter, and $q$ is the deceleration parameter. By computing the values of $r$ and $s$, one can analyze the cosmic characteristics of DE models without relying on specific theoretical frameworks. Certain values of the statefinder pair have well-known interpretations in the context of standard DE models. For example, $\{r,s\} = \{1,0\}$ corresponds to the $\Lambda$CDM model, which is the standard cosmological model with a cosmological constant. Similarly, $\{r,s\} = \{1,1\}$ corresponds to the standard cold dark matter (SCDM) model in a Friedmann-Lemaître-Robertson-Walker (FLRW) Universe. The region with $\{r,s\}$ values ranging from $(-\infty, \infty)$ corresponds to the Einstein static Universe. In the $r$-$s$ plane, different regions represent distinct types of DE models. For example, when $s > 0$, it indicates quintessence-like models, which involve a scalar field responsible for DE. Conversely, when $s < 0$, it indicates phantom-like models, characterized by the presence of a scalar field with negative kinetic energy. Deviations from the standard $\{r,s\} = \{1,0\}$ values can signify an evolutionary process from phantom-like to quintessence-like behavior. Moreover, specific combinations of the deceleration parameter $q$ and the statefinder parameter $r$ are associated with well-known models. For instance, $\{q,r\} = \{-1,1\}$ is linked to the $\Lambda$CDM model, which describes a Universe dominated by dark energy in the form of a cosmological constant. Another example is $\{q,r\} = \{0.5,1\}$, corresponding to the SCDM model, which assumes a Universe dominated by non-relativistic matter without dark energy.

\begin{figure}[!htb]
   \begin{minipage}{0.49\textwidth}
     \centering
   \includegraphics[scale=0.42]{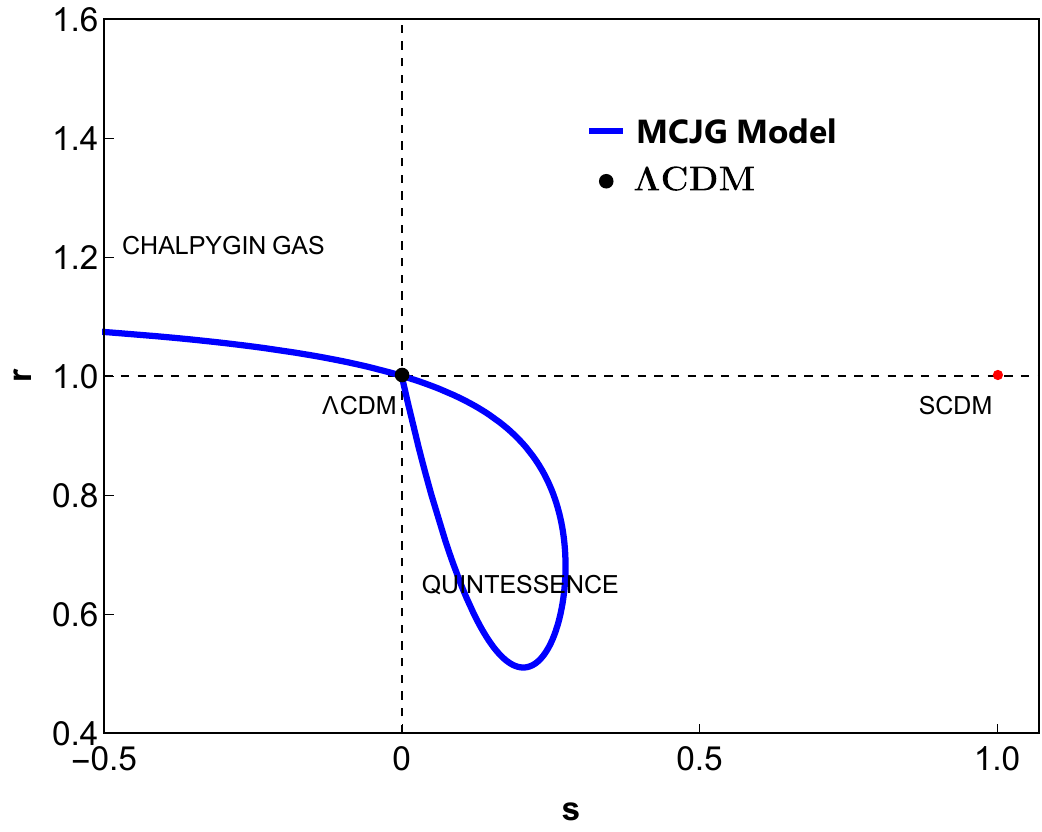}
\caption{This figure shows $\{s, r\}$ plots for MCJG Model .}\label{rs1}
   \end{minipage}\hfill
   \begin{minipage}{0.49\textwidth}
     \centering
    \includegraphics[scale=0.42]{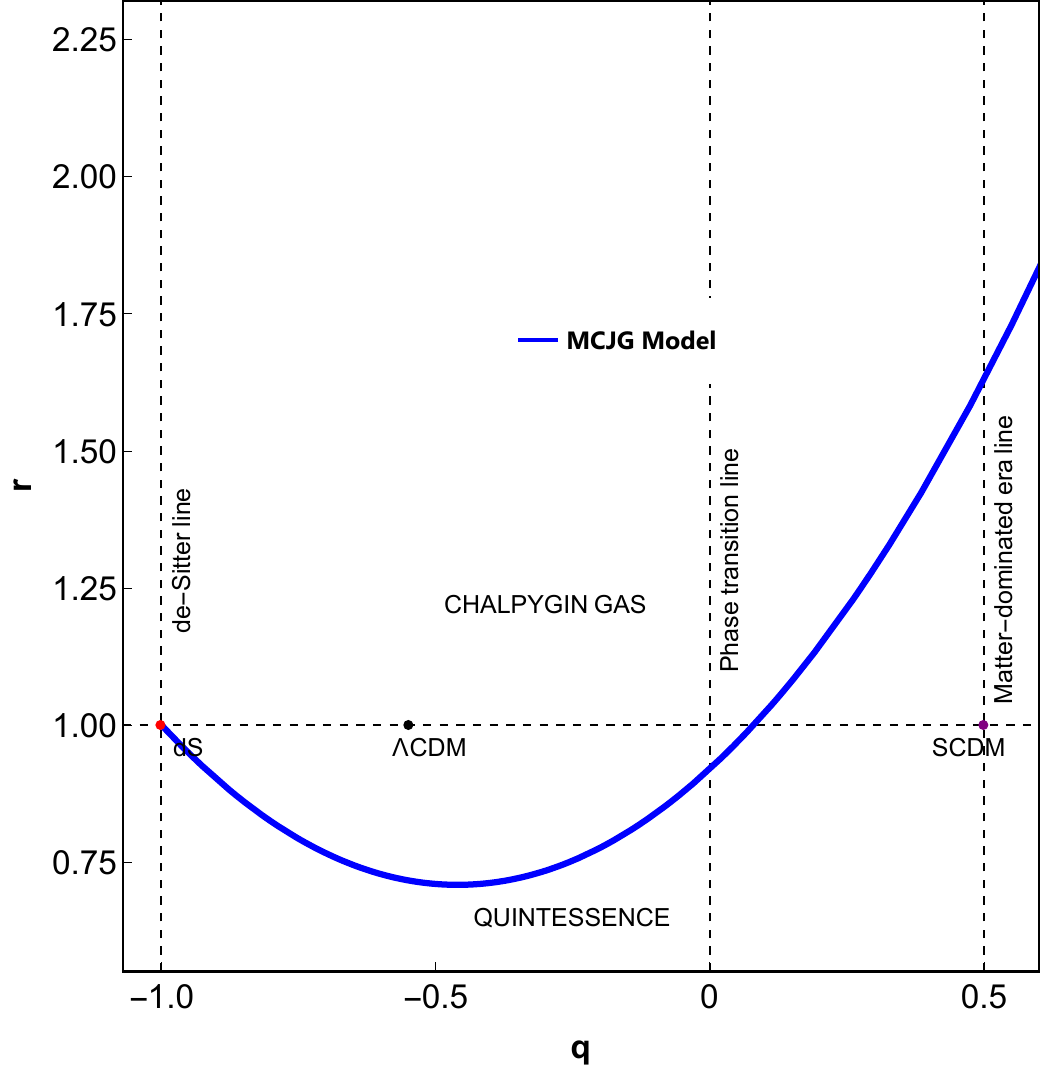}
\caption{This figure shows $\{q, r\}$ plots for MCJG Model.}
\label{rq1}
   \end{minipage}
\end{figure}
\begin{figure}[!htb]
   \begin{minipage}{0.49\textwidth}
     \centering
   \includegraphics[scale=0.42]{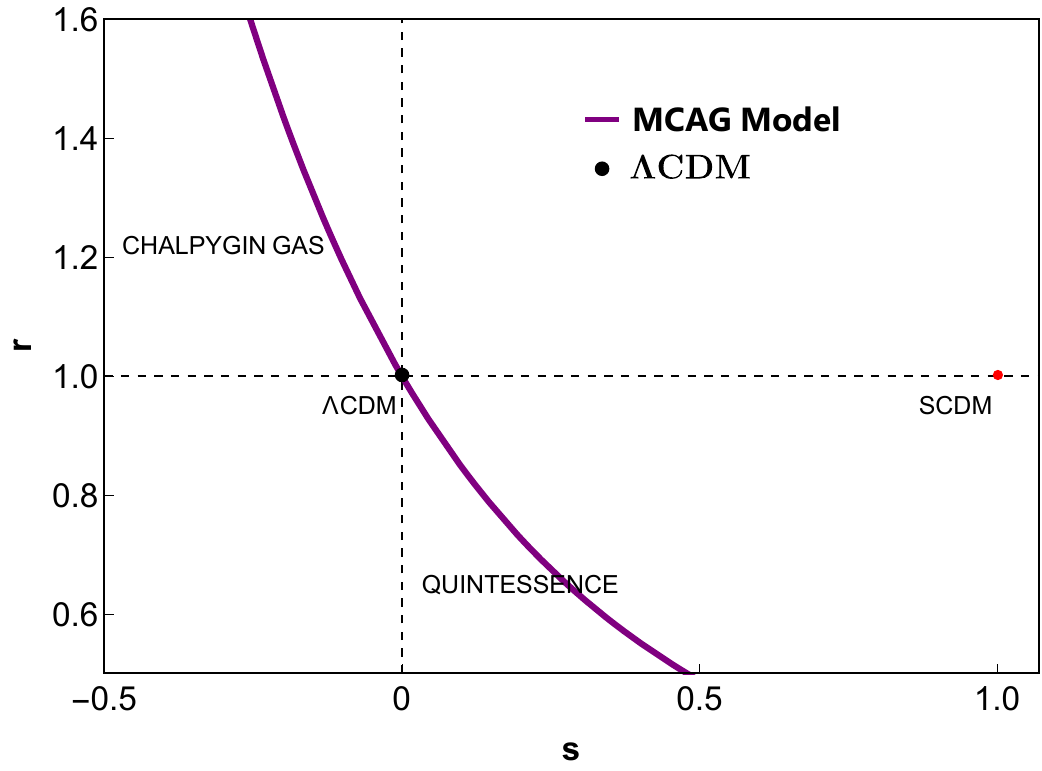}
\caption{This figure shows $\{s, r\}$ plots for MCAG Model .}\label{rs2}
   \end{minipage}\hfill
   \begin{minipage}{0.49\textwidth}
     \centering
    \includegraphics[scale=0.42]{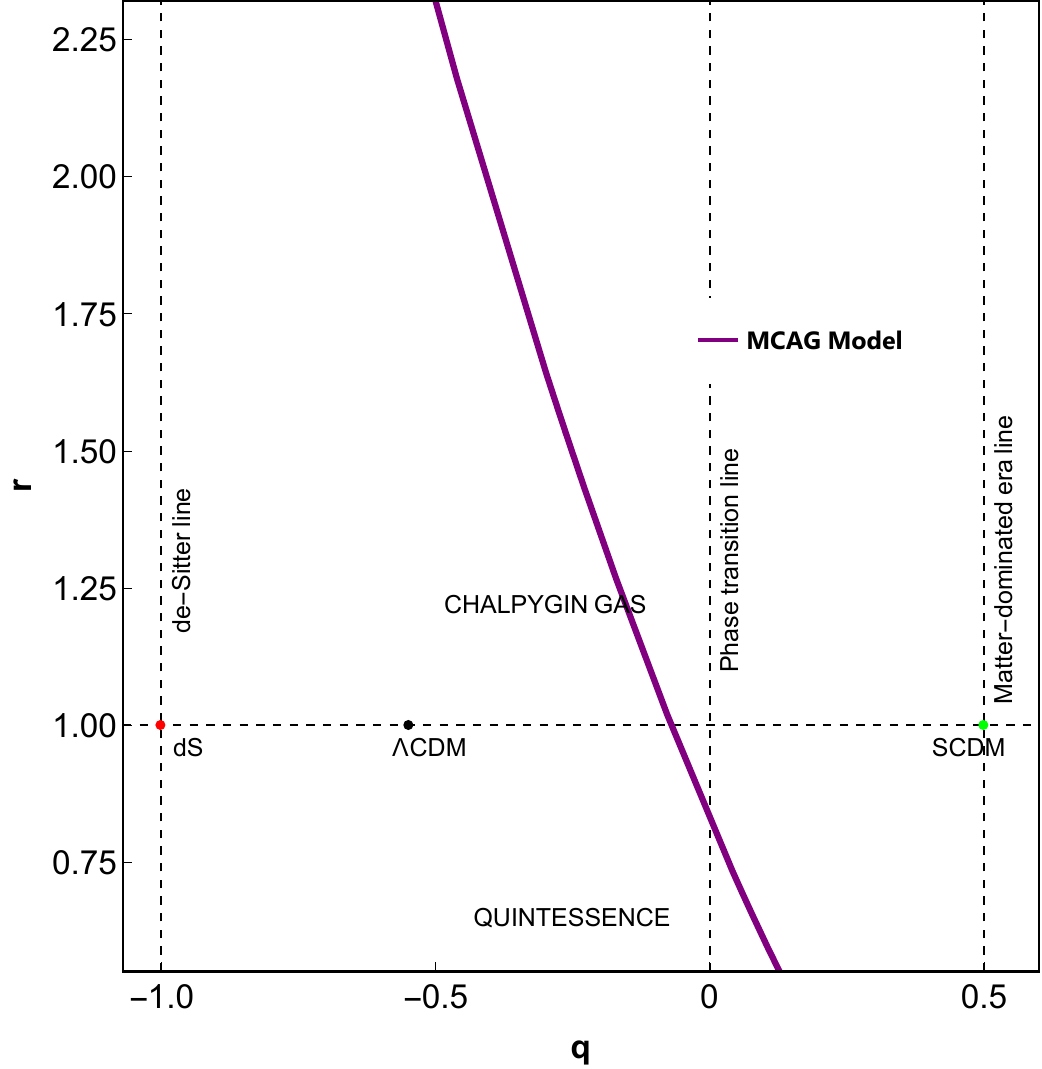}
\caption{This figure shows $\{q, r\}$ plots for MCAG Model.}
\label{rq2}
   \end{minipage}
\end{figure}

\section{Diagnostic Analysis}\label{secVIII}
The $Om(z)$ diagnostic is a valuable tool introduced in the literature ( \cite{Om1,Om2,Om3,Om4} ) that serves as an effective null test for the $\Lambda$CDM (Lambda Cold Dark Matter) model. This diagnostic, similar to the statefinder diagnostic, offers a viable approach for differentiating various dark energy (DE) models from the standard $\Lambda$CDM model by examining the slope variation of $Om(z)$. The  diagnostic provides insights into the nature of dark energy by characterizing its behavior and distinguishing between different regimes. The diagnostic is based on the analysis of the $Om(z)$ function, where $z$ represents the redshift, in a flat Universe. By studying the variation of $Om(z)$ with respect to redshift, distinct features and properties of the underlying dark energy model can be identified. A negative slope in the $Om(z)$ diagnostic corresponds to a phantom nature of dark energy. This suggests that the equation of state of dark energy is less than -1, indicating an accelerated expansion with even stronger dynamics compared to the $\Lambda$CDM model. Conversely, a positive slope in the $Om(z)$ diagnostic implies a quintessence regime for dark energy. In this case, the equation of state is greater than -1, indicating a slower rate of expansion compared to the phantom scenario. Furthermore, a constant slope of the diagnostic curve with respect to redshift corresponds to the $\Lambda$CDM model. This indicates that the dark energy component behaves like a cosmological constant throughout cosmic history, providing a good fit to observational data. The $Om(z)$ diagnostic thus enables us to discern different dark energy models from the $\Lambda$CDM model based on the characteristic slope variation of $Om(z)$. By analyzing observational data and comparing the slope of $Om(z)$, researchers can gain valuable insights into the nature of dark energy and the validity of alternative models. \\\\

\begin{equation}
\operatorname{Om}(z) = \frac{\left(\frac{H(z)}{H_0}\right)^2-1}{(1+z)^3-1}.
\end{equation}

\begin{figure}[!htb]
   \begin{minipage}{0.49\textwidth}
     \centering
   \includegraphics[scale=0.42]{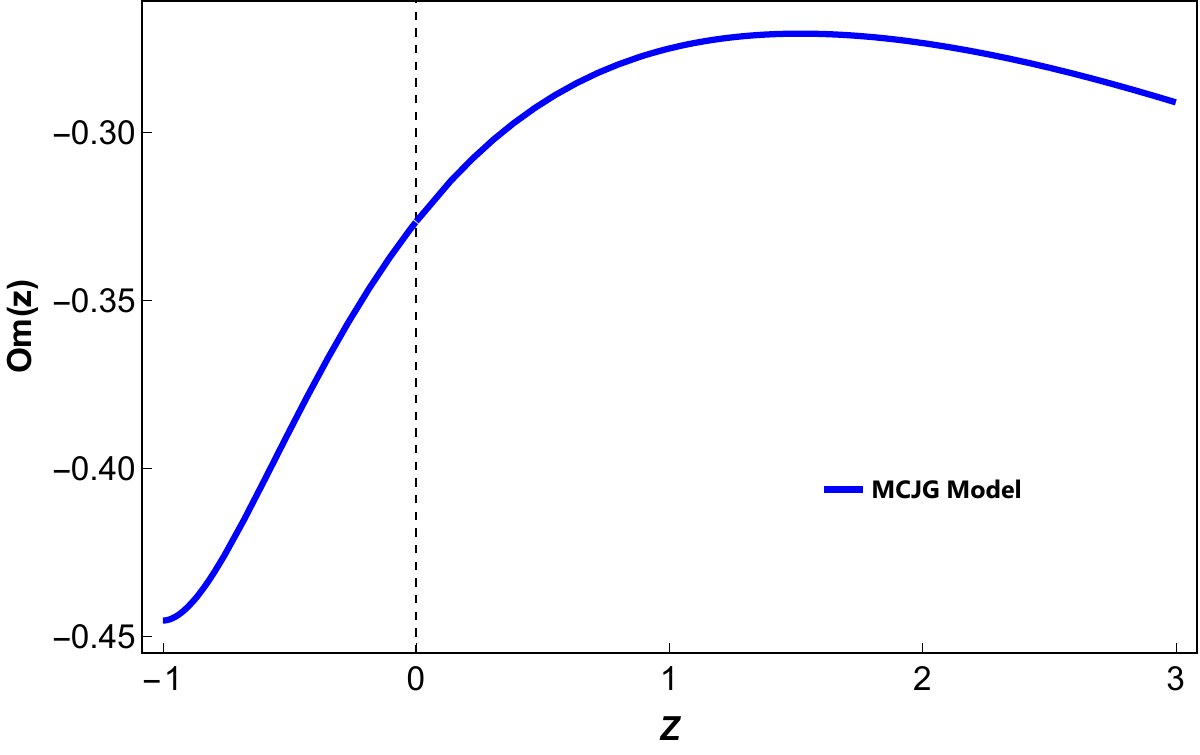}
\caption{This figure shows the $Om(z)$ with respect to redshift for MCJG Model.}\label{Om1}
   \end{minipage}\hfill
   \begin{minipage}{0.49\textwidth}
     \centering
    \includegraphics[scale=0.42]{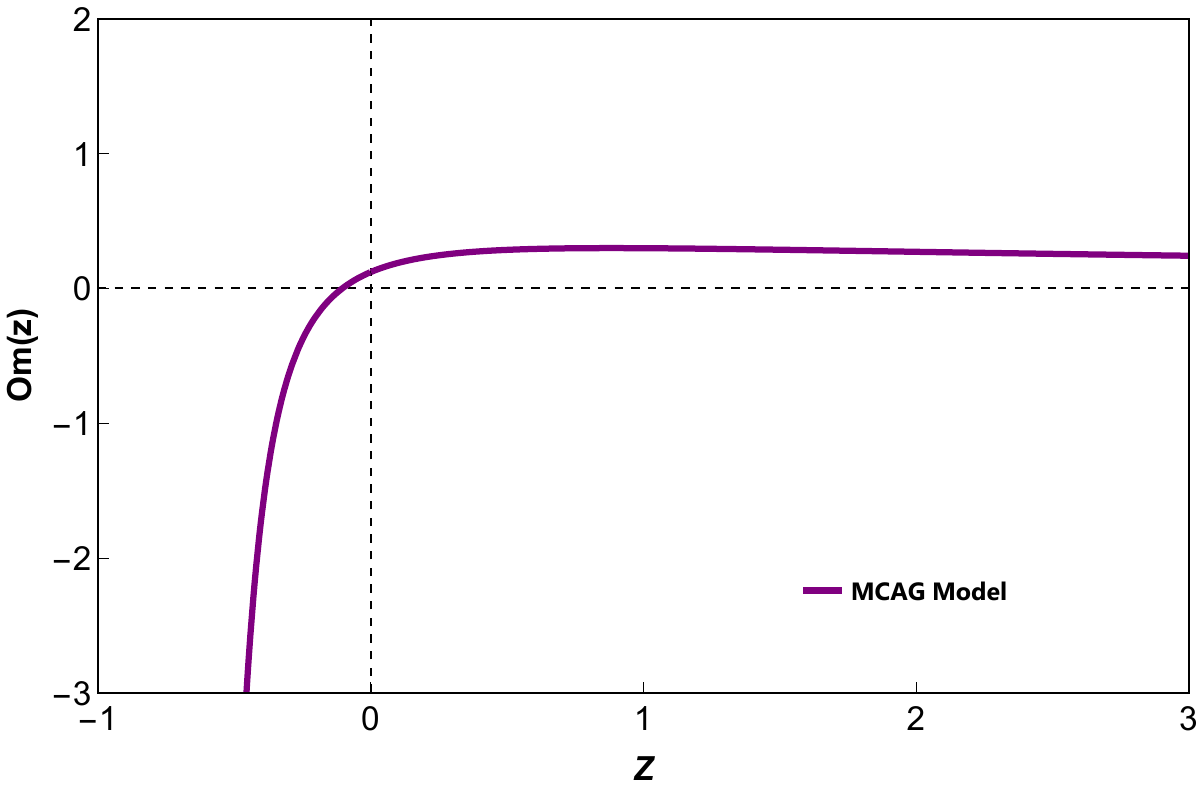}
\caption{This figure shows the $Om(z)$ with respect to redshift for MCAG Model.}\label{Om2}
   \end{minipage}
\end{figure}

\section{Information Criteria}\label{sec9}
To assess the viability of any model, a thorough understanding of information criteria (IC) is essential. The Akaike Information Criterion (AIC) \cite{AIC1,AIC2,AIC3,AIC4,AIC5} is commonly employed as a general IC tool. The AIC serves as an approximate measure of the Kullback-Leibler information divergence and asymptotically provides an unbiased estimator of this divergence. The AIC for Gaussian estimation is given by $\text{AIC}=-2\ln ({\cal
L}_{\text{max}})+2\kappa+\frac{2\kappa(\kappa+1)}{N-\kappa-1}$, where ${\cal L}_{\text{max}}$ denotes the maximum likelihood function, $\kappa$ represents the total number of free parameters in the model, and $N$ is the total number of data points used. Since the assumption is often $N\gg 1$ for the models considered, the formula simplifies to the original AIC form, $\text{AIC}=-2\ln ({\cal
L}_{\text{max}})+2\kappa$. When comparing multiple models, the differences in IC values can be quantified as $\triangle\text{AIC}=\text{AIC}_{\text{model}}-\text{AIC}_{\text{min}}=\triangle\chi^{2}_{\text{min}}+2\triangle\kappa$. The range of $\triangle\text{AIC}$ that is considered more favorable is $(0,2)$. The lower favorable range of $\triangle\text{AIC}$ is $(4,7)$, while values exceeding $\triangle\text{AIC}>10$ provide less support for the model. In addition to the AIC, the Bayesian information criterion ($BIC$) \cite{BIC,BIC2,BIC3} also contributes to model selection. Like the AIC, the $BIC$ considers the trade-off between goodness-of-fit and model complexity. However, the $BIC$ incorporates a stronger penalty for models with more parameters. This promotes the selection of simpler models, helping to guard against overfitting. In the context of $\Delta BIC$, similar principles apply as with $\Delta AIC$. A lower $\Delta BIC$ indicates stronger support for a particular model, and the magnitude of the difference is informative\\\\

\begin{table}[H]
\begin{center}
\begin{tabular}{|c|c|c|c|c|c|c|}  
\hline 
Model& $ \chi_{min}^{2} $& $ \chi_{red}^{2} $ &$ AIC $ & $ \Delta AIC $ &$ BIC $ & $ \Delta BIC $\\ \hline 
$ \Lambda$CDM & 2357.562 & 0.9837 & 2363.56 & 0 & 2379.08 & 0\\ \hline 
MCJG Model & 2342.767 & 0.9634 & 2364.77 & 1.205 & 2421.68 & 42.5967 \\ \hline 
\end{tabular}
\caption{Summary of ${\chi_{\text{tot}}^2}^{min} $, $\chi_{\text {red }}^2$, $A I C$, $\Delta A I C$, $B I C$, $\Delta B I C$}
\label{table3}
\end{center}
\end{table}

\begin{table}[H]
\begin{center}
\begin{tabular}{|c|c|c|c|c|c|c|}  
\hline 
Model& $ \chi_{min}^{2} $& $ \chi_{red}^{2} $ &$ AIC $ & $ \Delta AIC $ &$ BIC $ & $ \Delta BIC $\\ \hline 
$ \Lambda$CDM & 2558.341 & 0.9928 & 2564.34 & 0 & 2579.86 & 0\\ \hline 
MCAG Model & 2541.665 & 0.9919 & 2565.67 & 1.324 & 2627.75 & 47.8896 \\ \hline 
\end{tabular}
\caption{Summary of ${\chi_{\text{tot}}^2}^{min} $, $\chi_{\text {red }}^2$, $A I C$, $\Delta A I C$, $B I C$, $\Delta B I C$}
\label{table4}
\end{center}
\end{table}

\section{Results}\label{secIX}
\paragraph{deceleration parameter}
The comparison of the redshift dependence of the deceleration parameter for MCJG Model and MCJG Model with respect to the $\Lambda$CDM model is presented in Fig:- \ref{q(z) Model 2} and Fig:- \ref{q(z) Model 1}, respectively. In the considered redshift range of $z \in (-1,10)$, all three models exhibit similar behavior in the evolution of the deceleration parameter. However, in the MCJG Model , noticeable deviations from the $\Lambda$CDM model can be observed at high redshifts beyond $z = 4$. This indicates that the MCJG Model may introduce differences in the expansion dynamics compared to the standard $\Lambda$CDM model. The numerical values of the transition redshift, denoted as $z_{tr}$, which marks the transition from a decelerating phase to an accelerating phase, are relatively close for all three models, including the $\Lambda$CDM model. This implies that, in terms of the deceleration parameter, the models share similar characteristics regarding the shift from deceleration to acceleration in cosmic expansion. Finally, it is noteworthy that all three models exhibit a de Sitter phase characterized by a deceleration parameter of $q = -1$. The de Sitter phase corresponds to a phase of accelerated expansion driven by a cosmological constant or a similar dark energy component. These findings suggest that while the MCJG Model and the MCAJ Model generally resemble the behavior of the $\Lambda$CDM model in terms of the deceleration parameter, deviations become apparent at high redshifts in the MCJG Model. However, all models share similarities in terms of the transition redshift $z_{tr}$ and end in a de Sitter phase with $q = -1$, indicating accelerated expansion.\\\\

\paragraph{Jerk parameter}
The behavior of the jerk parameter, denoted as $j(z)$, for the (MCJG) Model and (MCAG) Model in comparison to the standard $\Lambda$CDM paradigm is depicted in Fig:- \ref{j(z) Model 1} and Fig:- \ref{j(z) Model 2}, respectively. These models provide alternative frameworks to explain the dynamics of the Universe and explore departures from the $\Lambda$CDM model. In Fig:- \ref{j(z) Model 1}, we observe that the predictions of the MCJG Model diverge significantly from the $\Lambda$CDM model at high redshifts. The MCJG Model exhibits a distinct behavior, showing a different trend in the evolution of cosmic acceleration. However, it is worth noting that at $z=-1$, both the MCJG Model and the $\Lambda$CDM model predict the same value for the jerk parameter. Similarly, in Fig:- \ref{j(z) Model 2}, we examine the behavior of the jerk parameter for the MCAG Model. This model also presents deviations from the $\Lambda$CDM model, particularly at higher redshifts. The MCAG Model showcases an opposite trend compared to the MCAG Model, suggesting a unique dynamic for cosmic acceleration. Remarkably, at $z=-1$, both the MCAG Model and the $\Lambda$CDM model yield identical values for the jerk parameter.\\\\

\paragraph{Snap parameter}
The behavior of the snap parameter, represented as $s(z)$, for the MCJG Model and MCAG Model, in comparison to the $\Lambda$CDM model, is depicted in Figure \ref{s(z) Model 1} and Figure \ref{s(z) Model 2}, respectively. These figures clearly illustrate significant deviations between the alternative models and the $\Lambda$CDM model, particularly at higher redshifts. In the context of cosmology, the snap parameter, $s(z)$, characterizes the rate of change of the acceleration of the Universe with respect to redshift. It is an important quantity for discerning the dynamic nature of dark energy.  Analyzing Figure \ref{s(z) Model 1}, we observe that the MCJG Model exhibits discrepancies compared to the $\Lambda$CDM model, with the differences becoming less pronounced at lower redshifts. This suggests that the MCJG Model aligns more closely with the $\Lambda$CDM model as redshift decreases, indicating a better agreement in terms of the snap parameter. On the other hand, Figure \ref{s(z) Model 2} illustrates that the MCAG Model displays a similar behavior to the $\Lambda$CDM model. The snap parameter for the MCAG Model exhibits characteristics that are consistent with the $\Lambda$CDM model, implying a comparable agreement in terms of the dynamics of acceleration. These observations imply that the MCJG Model and MCAG Model have distinct behaviors in comparison to the $\Lambda$CDM model, as evidenced by the variations in the snap parameter. While the MCJG Model converges towards the $\Lambda$CDM model at lower redshifts, the MCAG Model demonstrates a level of agreement with the $\Lambda$CDM model throughout the redshift range.\\\\

\paragraph{Statefinder diagnostic}
In Figure \ref{rs1}, the evolution of the statefinder pair $(r, s)$ for the MCJG Model is depicted. As the evolution progresses, the statefinder trajectory of the MCJG Model transitions from the quintessence region to the Chaplygin gas region at late times. This transition implies a change in the nature of dark energy, reflecting a shift in the dominant energy component driving the accelerated expansion of the Universe. It is worth noting that during this evolution, the MCJG Model bypasses the intermediate fixed point $(0, 1)$, which is associated with the standard $\Lambda$CDM model. Similarly, in Figure \ref{rs2}, the evolution of the statefinder pair $(r, s)$ for the MCAG Model is presented. The behavior of the MCAG Model is analogous to the MCJG Model as it undergoes a transition from the quintessence region to the Chaplygin gas region. This transition is again accompanied by bypassing the intermediate fixed point $(0, 1)$, which characterizes the $\Lambda$CDM model. In the context of cosmology, the quintessence region corresponds to a scenario where dark energy behaves like a dynamical scalar field, while the Chaplygin gas region represents a different regime where dark energy is modeled as a perfect fluid with an equation of state given by the Chaplygin gas equation. The trajectory of the statefinder pair $(r, s)$ provides valuable insights into the nature and evolution of dark energy within these alternative models. Fig \ref{rq1} provides additional insight into the behavior of the MCJG Model. As the evolution progresses, the trajectory of the MCJG Model undergoes a transition from the line corresponding to the Matter-dominated era. This region is characterized by a decelerating expansion due to the dominant matter content in the Universe. As the MCJG Model evolves further, it enters the quintessence region, which is indicated by the deviation from the Matter-dominated era line. In this region, the dominant energy component responsible for the cosmic acceleration is a quintessence scalar field with an equation of state greater than $-1$. This scalar field drives the accelerated expansion of the Universe. The trajectory of the MCJG Model ultimately converges towards the de Sitter point, denoted as $(-1, 1)$. At this point, the expansion of the Universe becomes asymptotically exponential, mimicking a cosmological constant (dark energy with a constant equation of state equal to -1). On the other hand, Fig \ref{rq2} illustrates the evolution of the statefinder pair $(q, r)$ for the MCAG Model. The trajectory of the MCAG Model undergoes a transition from the quintessence region, similar to the MCJG Model. However, in the case of the MCAG Model, the trajectory deviates towards the de Sitter point $(-1,1)$. This trajectory reveals the unique characteristics of the MCAG Model and its departure from the simple $\Lambda$CDM model.\\\\

\paragraph{$Om(z)$ diagnostic}
Figures \ref{Om1} and \ref{Om2} depict the behavior of the $Om(z)$ parameter with respect to redshift $z$ for the MCJG Model and MCAG Model, respectively. Analyzing Figure \ref{Om1}, we observe that for the MCJG Model, the $Om(z)$ values are smaller than $Om(z)$ for redshifts greater than zero ($z>0$). This indicates that the MCJG Model resides in the quintessence domain, where the expansion of the Universe is driven by a slowly-evolving quintessence field or dark energy component. As the redshift decreases, we notice a significant decrease in the $Om(z)$ value for the MCJG Model. Eventually, the value becomes negative, signifying a transition from the quintessence regime to the phantom region. The phantom region is characterized by an equation of state for dark energy that is less than $-1$, leading to an accelerated expansion with more dramatic dynamics than the standard $\Lambda$CDM model. Similarly, in Figure \ref{Om2} representing the MCAG Model, we observe that the $Om(z)$ values are also smaller than $Om(z)$ for $z>0$, placing the model within the quintessence domain. As the redshift decreases, the $Om(z)$ value exhibits a significant decrease and eventually becomes negative, indicating the transition from quintessence to the phantom region within the MCAG Model. The behavior observed in the transition from quintessence to phantom regions for both the MCJG and MCAG Models provides valuable insights into the cosmological dynamics and characteristics of these models. The change in the sign of $Om(z)$ signifies a shift from a slowly-evolving quintessence field to a more rapidly-evolving dark energy component, leading to an accelerated expansion of the Universe with distinct dynamics compared to the standard $\Lambda$CDM model.\\\\

\paragraph{Information Criteria}
In our analysis, we compared the MCJG and MCAG models with the well-established $\Lambda$CDM model. We provide the comprehensive comparison between the MCJG Model, the MCAG Model, and the $\Lambda$CDM Model in Table \ref{table3} and \ref{table4}. The comparison takes into account various statistical measures, including ${\chi_{\text{min}}^2}$, ${\chi_{\text{red}}^2}$, $AIC$, $\Delta AIC$, $BIC$, and $\Delta BIC$. The ${\chi_{\text{min}}^2}$ value for the MCJG Model is slightly lower than that of the $\Lambda$CDM Model, indicating a relatively better goodness of fit in terms of minimizing the total $\chi^2$. The ${\chi_{\text{red}}^2}$ value for the MCJG Model is lower than that of the $\Lambda$CDM Model, suggesting a slightly better fit when accounting for the degrees of freedom. The $AIC$ value for the MCJG Model is higher than that of the $\Lambda$CDM Model, indicating that the $\Lambda$CDM Model has a better balance between goodness of fit and model complexity according to AIC. The calculated $\Delta AIC$ of $1.205$ suggests that the MCJG Model is less supported compared to the $\Lambda$CDM Model. This value falls within the range of models that are less strongly supported according to AIC. The $BIC$ value for the MCJG Model is higher than that of the $\Lambda$CDM Model, supporting the notion that the $\Lambda$CDM Model is preferred in terms of model complexity and goodness of fit according to BIC. The calculated $\Delta BIC$ of $42.5967$ indicates that the $\Lambda$CDM Model is more favored compared to the MCJG Model. This substantial difference suggests strong evidence in favor of the $\Lambda$CDM Model according to BIC. The ${\chi_{\text{min}}^2}$ and ${\chi_{\text{red}}^2}$ values for the MCAG Model are lower than those of the $\Lambda$CDM Model, indicating a better goodness of fit in terms of minimizing the total $\chi^2$ and accounting for degrees of freedom. The $AIC$ value for the MCAG Model is higher than that of the $\Lambda$CDM Model, indicating that the $\Lambda$CDM Model has a better balance between goodness of fit and model complexity according to AIC. The calculated $\Delta AIC$ of $1.324$ suggests that the MCAG Model is less supported compared to the $\Lambda$CDM Model. This value falls within the range of models that are less strongly supported according to AIC. The $BIC$ value for the MCAG Model is higher than that of the $\Lambda$CDM Model, further supporting the notion that the $\Lambda$CDM Model is preferred in terms of model complexity and goodness of fit according to BIC. The calculated $\Delta BIC$ of $47.8896$ indicates that the $\Lambda$CDM Model is more favored compared to the MCAG Model. This substantial difference suggests strong evidence in favor of the $\Lambda$CDM Model according to BIC.

\section{Conclusion}

In this article, we consider a non-flat FLRW model of the Universe filled with radiation, dark matter and dark energy in the context of $f(T)$ gravity, which was proposed based on the corrections of the teleparallel equivalent of General Relativity (TEGR). It's 2nd order equations of motion makes it simpler than $f(R)$ gravity. Here we choose Modified Chaplygin-Jacobi gas (MCJG) and modified Chaplygin-Abel gas (MCAG) to play the role of dark energy. Also here we choose the power law form as viable $f(T)$ gravity model amongst various feasible $f(T)$ gravity models proposed in literature as this form  exhibits both matter dominated and radiation dominated era and successfully confronts the cosmological test with observational data. Here we obtain the Constraints on the model parameters for DE as well as the $f(T)$ gravity models by fitting recent astronomical data. Exploiting the observational data coming from H(z), Pantheon, Gamma Ray Bursts, quasars and BAO measurements, we have determined the best fit values of the model parameters. We present the reduced Hubble parameter in terms of observable parameters like $\Omega_{r0}, \Omega_{m0}, \Omega_{k0}, \Omega_{CJ0}, \Omega_{CA0}, H_0$ for both the considered dark energy models. In the figures \ref{MCJG} and \ref{MCAG}, we have drawn the contour plots of the MCMC confidence contours at 1$\sigma$ and 2$\sigma$ of MCJG model and  MCAG model respectively. The best fit values of the parameters of MCJG model and  MCAG model have been obtained by $\chi^{2}$ minimum test as
(i) $H_{0}$ = 69.158620,
$\Omega_{\mathrm{m0}}$ = 0.190482,
$\Omega_{\mathrm{k0}}$ = 0.001891,
$\Omega_{\mathrm{cjo}}$ = 0.001891,
$\alpha$ = 1.000928 
$\beta$ = 1.000928, 
$A_{s}$ = -0.929042,
K = 0.402158,
$\gamma$=1.000928.
(ii) $H_{0}$ = 69.158620,
$\Omega_{\mathrm{m0}}$ = 0.190482
,$\Omega_{\mathrm{k0}}$ = 0.001891,
$\Omega_{\mathrm{cjo}}$ = 0.001891,
$\alpha$=1.000928,
$\beta$=1.000928, 
$A_{s}$=-0.929042, 
K=0.402158,
$\gamma$=1.000928. Contrasting the predicted values of the considered cosmological models with the Hubble 36 observational dataset and $\Lambda$CDM paradim, we have seen the models adequetly matches the Hubble and the Pantheon dataset. In conclusion, our analysis of the MCJG Model and MCAG Model in comparison to the $\Lambda$CDM model has provided valuable insights into the dynamics of cosmic expansion and the nature of dark energy. Through various diagnostic tools, including the deceleration parameter, jerk parameter, snap parameter, and statefinder diagnostic, we have examined the behavior of these alternative models and their deviations from the standard model. The comparison of the deceleration parameter revealed that while all models share similarities in terms of the transition from deceleration to acceleration, noticeable differences were observed at high redshifts in the MCJG Model. This suggests that the MCJG Model introduces variations in the expansion dynamics compared to the $\Lambda$CDM model. However, all models ultimately converge to a de Sitter phase with a deceleration parameter of $q = -1$, indicating accelerated expansion driven by a cosmological constant or a similar dark energy component. The analysis of the jerk parameter showcased distinct behaviors of the MCJG Model and MCAG Model compared to the $\Lambda$CDM model, particularly at higher redshifts. These models exhibited different trends in cosmic acceleration, highlighting their departure from the standard model. However, at $z=-1$, both alternative models aligned with the $\Lambda$CDM model, suggesting similarities in the acceleration dynamics at that specific redshift. The examination of the snap parameter emphasized the differences between the MCJG Model and the $\Lambda$CDM model, with a closer agreement observed at lower redshifts. In contrast, the MCAG Model displayed behavior consistent with the $\Lambda$CDM model throughout the entire redshift range. These findings underscore the unique characteristics of these models in terms of their acceleration dynamics. Furthermore, the statefinder diagnostic provided insights into the nature of dark energy within the MCJG Model and MCAG Model. The transitions from the quintessence region to the Chaplygin gas region indicated changes in the dominant energy component responsible for cosmic acceleration. These transitions were accompanied by deviations from the intermediate fixed point associated with the $\Lambda$CDM model. The statefinder trajectories shed light on the evolution and behavior of dark energy within these alternative models. Our analysis highlights the distinct behaviors and deviations of the MCJG Model and MCAG Model from the $\Lambda$CDM model. These alternative models offer valuable insights into the dynamics of cosmic expansion and the properties of dark energy. While the MCJG Model exhibited notable differences at high redshifts, both models ultimately converged to a de Sitter phase characterized by accelerated expansion. The comparison of MCJG and MCAG with the $\Lambda$CDM model using $\Delta$AIC and $\Delta BIC$ values indicates moderate support for both models. While the $\Lambda$CDM model shows slightly better performance, further evidence from other observations is crucial to fully assess the viability of MCJG and MCAG in explaining dark energy and the universe's expansion. Our study has contributed to the ongoing quest for a deeper understanding of dark energy and its potential role in the cosmic acceleration. While the $\Lambda$CDM model remains a robust and successful framework for explaining various cosmological observations, exploring alternative models like MCJG and MCAG is vital for advancing our comprehension of the universe's fundamental properties. Our results open avenues for future research and inspire further investigations into the intriguing phenomena shaping the evolution of our cosmos
\section*{Conflict of Interest}
The authors declare that they have no known competing financial interests or personal relationships that could have appeared to influence the work reported in this paper.
\section*{Data Availability Statement}
This manuscript has no associated data or the data will not be deposited. [Authors comment: All data generated or analysed during this study are included in this published article [and its supplementary information files]

\bibliographystyle{elsarticle-num}
\bibliography{mybib}

\end{document}